\documentclass[prb,twocolumn,superscriptaddress,showpacs]{revtex4}
\usepackage{graphicx,bm,amssymb,amsmath}

\newcommand{\Eq}[1]{Eq.~\eqref{#1}}
\newcommand{\eq}[1]{\eqref{#1}}

\newcommand{\Sec}[1]{Section~\ref{#1}}
\newcommand{\Fig}[1]{Fig.~\ref{#1}}

\newcommand{\nn}{\nonumber}

\newcommand{\pdag}{{\phantom{\dagger}}}
\newcommand{\pprime}{{\phantom{\prime}}}

\newcommand{\bra}[1]{\left\langle{#1}\right\rvert}
\newcommand{\ket}[1]{\left\lvert{#1}\right\rangle}

\DeclareMathOperator{\sign}{sgn}

\DeclareMathOperator{\im}{Im}
\DeclareMathOperator{\re}{Re}
\DeclareMathOperator{\tsum}{\textstyle\sum}

\newcommand{\PRL}[3]{Phys. Rev. Lett.~\textbf{#1}, #2 (#3)}
\newcommand{\PRB}[3]{Phys. Rev. B~\textbf{#1}, #2 (#3)}

\newcommand{\RMP}[3]{Rev. Mod. Phys.~\textbf{#1}, #2 (#3)}

\newcommand{\Nature}[3]{Nature~\textbf{#1}, #2 (#3)}
\newcommand{\JETP}[3]{Sov. Phys. JETP~\textbf{#1}, #2 (#3)}
\newcommand{\ZhETF}[3]{Zh. Eksp. Teor. Fiz.~\textbf{#1}, #2 (#3)}

\newcommand{\NPB}[3]{Nucl. Phys. B~\textbf{#1}, #2 (#3)}
\newcommand{\JMP}[3]{J. Math. Phys.~\textbf{#1}, #2 (#3)}
\newcommand{\JPA}[3]{J. Phys. A~\textbf{#1}, #2 (#3)}
\newcommand{\SSC}[3]{Sol. State Commun.~\textbf{#1}, #2 (#3)}

\begin{document}

\title{
Fermi-Luttinger liquid: spectral function of interacting
one-dimensional fermions
}

\author{M. Khodas}
\affiliation{William I. Fine Theoretical Physics Institute and
School of Physics and Astronomy, University of Minnesota,
Minneapolis, MN 55455}
\author{M. Pustilnik}
\affiliation{School of Physics, Georgia Institute of Technology,
Atlanta, GA 30332}
\author{A. Kamenev}
\affiliation{ School of Physics and Astronomy, University of
Minnesota, Minneapolis, MN 55455}
\author{L.I. Glazman}
\affiliation{William I. Fine Theoretical Physics Institute and
School of Physics and Astronomy, University of Minnesota,
Minneapolis, MN 55455}

\begin{abstract}
We evaluate the spectral function of interacting fermions in one
dimension. Contrary to the Tomonaga-Luttinger model, our treatment
accounts for the nonlinearity of the free fermion spectrum. In a
striking departure from the Luttinger liquid theory, the spectrum
nonlinearity restores the main feature of the Fermi liquid: a Lorentzian peak
in the spectral function on the \textit{particle} mass-shell. At the
same time, the spectral function displays a power-law singularity on
the \textit{hole} mass-shell, similar to that in the Luttinger liquid.
\end{abstract}


\maketitle

\section{Introduction}
\label{Sec:Intro}

One-dimensional problems play a special role in quantum many-body
theory. In many cases, the reduced dimensionality affords one a
deeper understanding of the role of interactions in a many-body
system. Recent progress in experimental techniques has also
contributed to the increased attention paid to a variety of
one-dimensional (1D) fermionic and bosonic systems. Examples
include edge modes of the quantum Hall liquid~\cite{Chang1983},
carbon nanotubes~\cite{Deccer1998McEuen1999}, cleaved edge
semiconductor wires~\cite{Yacoby}, antiferromagnetic spin
chains\cite{Nagler}, and cold atoms in 1D optical
traps~\cite{ColdAtoms}. These developments catalyze the interest
to the fundamental theory of interacting 1D quantum
liquids~\cite{Drag2003,Wiegmann2005,Pustilnik2006,
Pereira2006,Teber2006,MP_CSM,Caux}.

The traditional framework for discussing 1D systems is provided
by the exactly solvable
Tomonaga-Luttinger  (TL) model\cite{TL1,TL2,TL3}. The crucial
simplification that makes the model solvable is the linearization
of the fermionic dispersion relation. It was understood early on\cite{TL2}
that a model with a linear spectrum is an idealized one.
The goal of this paper is to elucidate the influence of the dispersion
nonlinearity on the spectral function of 1D Fermi systems.

In the absence of interactions, the spectral function is given by
$A_p(\epsilon) = \delta(\epsilon - \xi_p)$, where $\xi_p$ is the
single-particle energy measured relative to the Fermi level
($\xi_p=p^2/2m-\epsilon_F$ for Galilean-invariant systems).
According to the Fermi liquid theory \cite{Nozieres},   weak
repulsive interactions merely broaden the peak in $A_p(\epsilon)$
to a Lorentzian,
\begin{equation}
A_p(\epsilon) \propto -
\im\frac{1}{\epsilon-\xi_p + i/2\tau_p},
\label{FL}
\end{equation}
The broadening originates in the finite decay rate $1/2\tau_p$ of the
Fermi liquid quasiparticles. The rate can be estimated with the help
of the Golden Rule,
\begin{equation}
\frac{1}{2\tau_p} \sim (D-1)(\nu
V)^2\,\frac{\xi_p^2}{\epsilon_F},
\label{selfenergy}
\end{equation}
where $\nu$ is $D$-dimensional density of states and $V$ is the
characteristic strength of the short-range repulsive interaction.
Indeed, quasiparticle relaxation occurs via real transitions resulting
in the excitation of particle-hole pairs. In dimensions $D>1$ already
a single pair produces a finite decay rate. The amplitude of such
process is proportional to $V$, hence the dimensionless factor
$(\nu V)^2$ in \Eq{selfenergy}. The factor $\xi_p^2$ accounts for
the corresponding phase space volume, i.e. the number of possibilities
a pair can be excited while obeying the energy and momentum
conservation laws.

In 1D the situation differs dramatically. Indeed, the Golden Rule
result \Eq{selfenergy} is identically zero for $D=1$. Moreover,
it can be shown\cite{DL} that in the framework of the TL
model~\cite{TL1,TL2,TL3}, the self-energy vanishes on the
mass-shell in {\em all} orders of the perturbation theory. In
fact, the TL Green function~\cite{DL,LP} assumes manifestly
non-Fermi liquid form. In the vicinity of the mass-shell,
$|\epsilon-\xi_k|\ll |\xi_k|$, it reads
\begin{equation}
G_k(\epsilon) \propto \left(
\frac{-\sign\xi_k}{\epsilon- \xi_k +i0} \right)^{1-\gamma_0^2}\, ,
\label{TLGreen}
\end{equation}
 where $k=p-p_F$ and $\xi_k=\pm \,vk$ is the
dispersion relation linearized near the right (left) Fermi point
$\pm\,p_F$. The corresponding spectral function of, say, the right-moving
excitations then takes the form (hereinafter we concentrate on
 the limit of zero temperature\cite{lehur})
\begin{equation}
A_k(\epsilon) \sim
\frac{\gamma_0^2}{|\epsilon-\xi_k|^{1-\gamma_0^2 }}\,\,
\theta\bigl[(\epsilon-\xi_k)\sign\xi_k\bigr].
\label{TL}
\end{equation}
The exponent $\gamma_0$ in Eqs. \eq{TLGreen} and \eq{TL}
characterizes the interaction  strength and in the lowest order is given by
\begin{equation}
\gamma_0 = \frac{1}{2}\,\nu\,(V_0-V_{2p_F}),
\label{gamma0}
\end{equation}
 where $\nu = (2\pi v)^{-1}$ is the density of
states in 1D and $V_k$ is the Fourier transform of a short-range
interaction potential.

Unlike the Fermi-liquid Lorentzian \Eq{FL}, the spectral function
\Eq{TL} exhibits a characteristic threshold behavior. The edges of
the spectral support coincide with the single-particle energies
$\xi_k$, and the spectral function displays  a power-law edge
singularity on the mass-shell $\epsilon\to\xi_k$. This singularity
is a hallmark of the Luttinger liquid~\cite{Haldane} behavior. The
particle-hole symmetry of the TL model implies that $A_k(\epsilon)
= A_{-k}(-\epsilon)$, hence the behavior of $A_k(\epsilon)$ in the
particle region of the spectrum $\epsilon>0$ is identical to that
in the hole region $\epsilon<0$.

The exact solvability of the TL model relies crucially on the
assumption of strictly linear dispersion relation. The purpose of this
paper is to examine the effects of the dispersion nonlinearity on the
spectral function of 1D spinless fermions. Specifically, we
consider a nonlinear dispersion relation with a positive
curvature, and approximate the single-particle spectrum in the
vicinity of the right $(R)$ and left $(L)$ Fermi points by
\begin{equation}
\xi_k^{R/L} =
\pm\, vk + \frac{k^2}{2m} + \ldots,
\quad k = p\mp p_F.
\label{dispersion}
\end{equation}
The presence of a finite mass $m$ breaks the particle-hole symmetry of
the TL model and affects the spectral function in the particle and hole
regions of the spectrum in manifestly different ways.

 \begin{figure}[t]
\includegraphics[width=0.72\columnwidth]{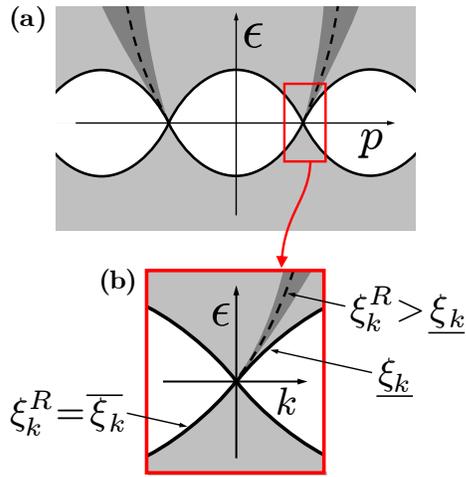}
\caption{
(a) Support of the spectral function in $(p, \epsilon)$-plane.
The parabola represents the mass-shell $(\epsilon=\xi_p)$. For
$\epsilon<0$, the mass-shell coincides with the edge of the
spectral support. For $\epsilon>0$, however, the mass-shell
(dashed lines) falls into a broad spectral continuum. In this
region, the spectral function has a peak of a finite width, which
is indicated schematically by dark gray.  (b) The
low-energy region near the right Fermi point $p = p_F$, with
$k=p-p_F$.
}
\label{Fig1}
\end{figure}

The effect of the dispersion nonlinearity on the \textit{particle}
region $\epsilon>0$ is the most dramatic. Rather than being the edge
of the spectral support, the mass-shell $\epsilon=\xi_k^R$ now falls
within a broader continuum, see \Fig{Fig1}. Consequently, the
edge singularity [cf. \Eq{TL}] on the mass-shell disappears and gets
replaced by a peak of a finite height.
Both the shape and the width of the peak appear to be rather different
from those in $D>1$, see Eqs.~\eq{FL} and \eq{selfenergy}.
For $\epsilon >0$ and $|\epsilon -\xi^R_k| \lesssim k^2\!/2m$ we found
\begin{widetext}
\begin{equation} A_k(\epsilon) \propto
\im
\left\lbrace
\frac{\mu_{2p_F+k}^2}{\gamma_k^2}\, \left(
\frac{-1}{\epsilon-\xi_k^R+i/2\tau_k} \right)^{1-\gamma_k^2}\!
- \,\, \frac{\mu_{k}^2}{\gamma_k^2} \, \left(
\frac{1}{\epsilon-\xi_k^R+i/2\tau_k} \right)^{1-\gamma_k^2}
\right\rbrace.
\label{result}
\end{equation}
\end{widetext}
The $k$-dependent exponent $\gamma_k$ is given by
\begin{equation}
\gamma_k^2 =
\frac{1}{4}\bigl(\mu_{2p_F+k}^2+\mu_k^2\bigr),
\label{gammak}
\end{equation}
where
\begin{equation}
\mu_k = \nu\left(V_0-V_k \right) \frac{2m v}{|k|}\,
\label{mu}
\end{equation}
is the exponent governing the power-law divergence of the dynamic
structure factor\cite{Pustilnik2006}. For  short-range
interactions $\mu_k\to 0$ when $k\to 0$, while
$\mu_{2p_F}=2\gamma_0$, hence $k\to 0$ limit of \Eq{gammak}
agrees with \Eq{gamma0}.

The main feature of \Eq{result} is the appearance of a finite
quasiparticle decay rate, $1/2\tau_k$. To the lowest nonvanishing
order in the interaction strength it is given by
\begin{equation}
\frac{1}{2\tau_k}
= C\bigl[\nu^2V_0(V_0-V_k)\bigr]^2\frac{(\xi_k^R)^4}{(mv^2)^3}\,,
\quad
C = \frac{3^3\pi}{5\cdot 2^9}\,\approx \,0.03\,.
\label{1dselfenergy}
\end{equation}
A finite decay rate emerges only in the fourth order in the
interaction strength~\cite{footnote}. This is because the minimal
relaxation process involves excitation of two particle-hole pairs on
the opposite branches of the spectrum. The factor $(\xi_k^R)^4$
in \Eq{1dselfenergy} accounts for the corresponding phase space volume.
Note also that $1/\tau_k\propto m^{-3}$ and vanishes in the limit
$m\to\infty$ taken at a fixed $v$, which corresponds to the TL model.

\begin{figure}[b]
\centerline{\includegraphics[width=0.75\columnwidth]{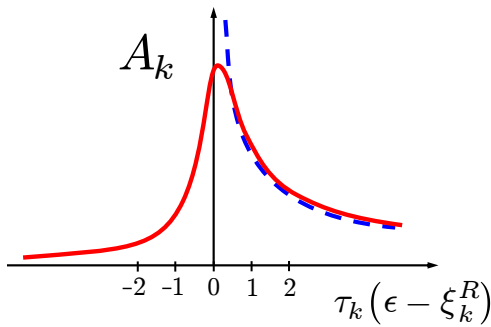}}
\caption{
Spectral function in the vicinity of the particle mass-shell
$\epsilon\approx\xi_k^R > 0$. The solid line is a plot of \Eq{result}
with $\mu_{2p_F+k} =1.1$, $\mu_k = 0.4$, and $\gamma_k$
given by \Eq{gammak}. For comparison, the dashed line corresponding
to the Luttinger liquid result \Eq{TL} is also shown.}
\label{Fig:SpecDens}
\end{figure}

Despite its small value compared to that in higher dimensions, the very
emergence of a finite quasiparticle relaxation rate in a 1D system is
a matter of fundamental significance. One may even wonder if it puts 1D
Fermi systems back in the realm of the conventional Fermi liquid
theory. Indeed, in a broad region $|\epsilon-\xi_k^R|\lesssim
(\tau_k\gamma_k^2)^{-1}$ around the mass-shell, the spectral function
\Eq{result} is essentially a Lorentzian with the width
$1/\tau_k\ll (\tau_k\gamma_k^2)^{-1}\ll \xi_k^R$, see
\Fig{Fig:SpecDens}. Restoring all the factors, we find for the
immediate vicinity of the particle mass-shell,
$|\epsilon-\xi_k^R|\lesssim (\tau_k\gamma_k^2)^{-1}$,
\begin{equation}
A_k(\epsilon) = \frac{1}{\pi}\,
\left(\frac{k}{mv}\right)^{\!\!3\gamma_0^2}\!
\left(\frac{m}{k^2\tau_k}\right)^{\gamma_k^2}\!\!
\frac{1/2\tau_k}{(\epsilon-\xi_k^R)^2+1/4\tau_k^2}\,.
\label{Lorentzian}
\end{equation}
Furthermore, for not too small momenta, such that  
\begin{equation}
\ln (k/p_F)\gtrsim -\,\frac{1}{\gamma_0^2}
\end{equation}
(this condition allows for $k\ll p_F$ at $\gamma_0\ll 1$) the
Lorentzian peak \eq{Lorentzian} carries most of the particle's
spectral weight, see \Sec{conclusions}.  This is the hallmark of the
Fermi liquid~\cite{Nozieres}.  The Luttinger liquid behavior is found
only at smaller momenta and sufficiently far away from the mass-shell,
at $|\epsilon-\xi_k^R|\gg (\tau_k\gamma_k^2)^{-1}$.

The dispersion nonlinearity has a strong effect on the spectral function
despite being irrelevant in the renormalization group (RG) sense:
it does not affect, for example, the power-law asymptote of
the local tunneling density of states at low energies.  However,
momentum-resolved measurements~\cite{Yacoby,Nagler}
and numerical simulations~\cite{Pereira2006} may reveal
the structure of the quasiparticle peak near the particle mass-shell.

In the hole region of the spectrum $\epsilon<0$, the positive dispersion
curvature does not smear the TL power-law singularity at the
mass-shell. The singularity is preserved because due to kinematic
constrains, the hole mass-shell remains to be the edge of the
spectral support (see the next section). The only effect of the
spectrum nonlinearity is the renormalization of the exponent in the
hole part of the spectral function [cf. \Eq{TL}],
\begin{equation}
 A_k(\epsilon) \propto
\frac{\gamma_k^2}{(\xi_k^R-\epsilon)^{1-\gamma_k^2 }}\,\,
\theta\bigl(\xi_k^R-\epsilon \bigr),
\label{Ahole}
 \end{equation}
where $\xi_k^R-\epsilon\ll k^2/2m$ and the exponent $\gamma_k^2$
is given by \Eq{gammak}. Note that the exponent is invariant upon
momentum inversion $k\leftrightarrow 2p_F+k$ and interpolates
smoothly between the two Fermi points; \Eq{Ahole} is applicable
along the entire hole mass-shell line, see \Fig{Fig1}. The
quadratic dependence of the exponent on the interaction strength
originates in the orthogonality catastrophe phenomenon~\cite{Anderson}.
Further away from the hole mass-shell the spectral function crosses
over to the TL one, \Eq{TL}.

We described above the behavior of the spectral function close to
the mass-shell. The regions of a finite spectral weight
are indicated by the shaded area in \Fig{Fig1}.
Apart from the hole mass-shell, all edges of the spectral support
are characterized by the power-law behavior of the spectral function with
momentum-dependent \textit{positive} exponents. This is easy to
foresee, if one notices that the states responsible for the
nonzero spectral weight far from the single-particle mass-shell
must involve several excitations. The main contributions to the
values of the exponents come from the constraints on the phase
space available for such excitations. The additional contribution
to the exponents is due to the interaction between the excited
particles and holes. This contribution is first-order in
the interaction potential and has the same origin as the the
exponents in the X-ray singularity phenomenon~\cite{Mahan}.
A detailed theory of the threshold behavior of the spectral function
is developed in \Sec{Sec:Edges}.

The rest of the paper is organized as follows: In
\Sec{Sec:Discussion} we present a qualitative analysis of the
problem. Perturbative calculation of $1/\tau_k$ to fourth order in the
interaction strength is carried out in detail in \Sec{Sec:PertTheor}. 
In \Sec{Sec:LogRenorm} we develop a strategy of summing up the 
leading logarithmic corrections while accounting for a finite $1/\tau_k$, 
and derive \Eq{result}. The behavior near the edges of the spectral 
support is discussed in \Sec{Sec:Edges}. We compare our findings 
with the results obtained for the exactly solvable Calogero-Sutherland 
model in \Sec{Sec:Calogero}. Finally, discussion and outlook are 
presented in \Sec{conclusions}.

\section{Qualitative considerations}
\label{Sec:Discussion}

In this section we discuss the boundaries (kinematic edges) of the
area in $(p,\epsilon)$-plane where the spectral function differs
from zero (see \Fig{Fig1}). We also provide a simple
qualitative framework for understanding the behavior of the
spectral function near the kinematic edges and in the vicinity of
the particle mass-shell.

\subsection{Support of the spectral function}

Using the Lehmann representation, we may express the particle
contribution to the spectral function as
\begin{equation}
A_k(\epsilon)
\propto
\sum_{\ket{f}}\bigl\lvert\bra{f}\psi_k^{R\dagger}\!\ket{0}\bigr\rvert^{2}\,
\delta\bigl(k-P_{\ket{f}}+P_{\ket{0}}\bigr)
\delta\bigl(\epsilon-E_{\ket{f}} + E_{\ket{0}}\bigr).
\label{2.1}
\end{equation}
Equation \eq{2.1} may be viewed as the probability  of tunneling
of a particle with a given momentum $p_F +k$ and energy
$\epsilon_F+\epsilon$ into a 1D system (to be definite, we
consider the right-movers $p_F+k >0$). The initial state
of the transition in \Eq{2.1} is the ground state of the system
$\ket{0}$, and the final state is $\ket{f}$, with
$P_{\ket{0}}$, $P_{\ket{f}}$ and $E_{\ket{0}}$,
$E_{\ket{f}}$ being the corresponding momenta (relative to $p_F$)
and energies (relative to $\epsilon_F$), respectively.

In the absence of interactions, the only possible final state is that
with a single right-moving particle added to the ground
state. Because of the momentum conservation the added particle
must have momentum $k$ (hereinafter the single-particle momenta
are measured relative to the respective Fermi points), $\ket{f}
=\psi_{k}^{R\dagger}\ket{0}$. Equation~(\ref{2.1}) then yields
$A_k\propto\theta(k)\,\delta(\epsilon-\xi_k^R)$. With interactions
present, states $\ket{f}$ may contain, in addition, a number of
particle-hole pairs. This allows $A_k(\epsilon)$ to be finite
away from the single-particle mass-shell $\epsilon = \xi_k^R$.
Still, there are regions in $(p,\epsilon)$-plane where
$A_k(\epsilon)=0$ due to kinematic constrains. To simplify the
analysis of  these constraints we focus on the low energies,
$|\epsilon|\ll\epsilon_F$, and small momenta $|k|\ll k_F$.

\begin{figure}[h]
\includegraphics[width=0.75\columnwidth]{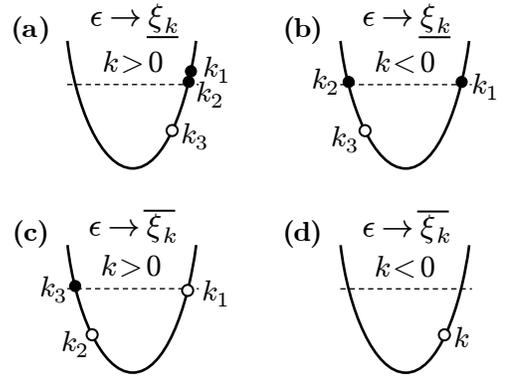}
\caption{
Final states $\ket{f}$, see Eqs.~\eq{2.1} and \eq{2.5},
corresponding to the boundaries of the regions
where $A_k(\epsilon)>0$. The black (white) circles stand for extra
particles (holes) added to the ground state; the dashed line indicates
the Fermi level. States (a) and (b) correspond to the particle edge
($\epsilon>0$); states (c) and (d) correspond to the hole edge.
}
\label{Fig_states}
\end{figure}

The simplest low-energy final state, beyond the single-particle one,
is the state containing one additional particle-hole pair.  Given that
the total momentum (relative to $p_F$) is small, the additional pair
must have a small momentum too. Therefore, it has to be located in the
vicinity of either the right or the left Fermi point. These states,
see Figs.~\ref{Fig_states}(a) and \ref{Fig_states}(b), have the form
\begin{equation}
\ket{f} =
\psi_{k_1}^{R\dagger}\psi_{k_2}^{R\dagger}\psi_{k_3}^{R\pdag}\!\ket{0},
\quad
k_1>0\,,~k_2>0\,,~k_3<0
\label{state1}
\end{equation}
and
\begin{equation}
\ket{f} =
\psi_{k_1}^{R\dagger}\psi_{k_2}^{L\dagger}\psi_{k_3}^{L\pdag}\!\ket{0},
\quad
k_1>0\,,~k_2<0\,,~k_3>0,
\label{state2}
\end{equation}
respectively.
The state \eq{state1} has energy
\begin{equation}
E_{\ket{f}}-E_{\ket{0}}
= v(k_1+k_2-k_3) + \frac{1}{2m}\left(k_1^2+k_2^2-k_3^2\right).
\label{2.2}
\end{equation}
Taking into account the momentum conservation $k_1+k_2-k_3=k$, we
find
\begin{equation}
E_{\ket{f}}-E_{\ket{0}}
= vk+ \frac{1}{2m}\left[k_1^2+k_2^2-(k_1+k_2-k)^2\right].
\label{2.3}
\end{equation}
The constraints on $k_1$, $k_2$, and $k_3$ in \Eq{state1}
guarantee that $k>0$. At a given $k$, the smallest possible value
of the excitation energy $E_{\ket{f}}-E_{\ket{0}}$ is reached
at $k_1=k_2=0$, $k_3=-k$.

Similar consideration for the state \eq{state2} yields
(with momentum conservation $k_1+k_2-k_3 =k$ taken into account)
\begin{equation}
E_{\ket{f}}-E_{\ket{0}}
=- vk + 2vk_1- \frac{k^2}{2m} + \frac{k}{m}(k_1+ k_2) - \frac{k_1 k_2}{m}\,.
\label{2.4}
\end{equation}
 At $k<0$, the lowest energy is reached at $k_1=k_2=0$,
$k_3=-k$.

An important observation following from the above analysis is that the
lowest possible excitation energy corresponds to the final states
with all particles at the Fermi level, and a \textit{single hole}
with the largest possible absolute value of the momentum. It is
easy to check that a final state with more than one particle-hole pair
excited from the ground state still has the lowest energy when
all the available momentum is ``carried'' by just a single hole.
Thus the final states \eq{state1} and \eq{state2} correspond to
the kinematic boundary $\underline{\xi_k\!}$ for the particle
($\epsilon>0$) part of the spectral function. Combining
Eqs.~\eq{2.3} and \eq{2.4} at $k_1=k_2=0$, we find
\begin{equation}
\underline{\xi_k\!} = v|k| - k^2\!/2m.
\label{xiunder}
\end{equation}
Although we considered here the small-$k$ domain only, \Eq{xiunder}
is valid for the entire region $-2p_F<k<2p_F$, see Fig~\ref{Fig1}.
Finding the spectral edges for higher momenta $|k|>2p_F$, however,
requires consideration of final states with more than one particle-hole
pair.

When considering the hole part of the spectrum, we again start
with the Lehmann representation,
\begin{equation}
A_k(\epsilon) \propto
\sum_{\ket{f}}\bigl\lvert\bra{f}\psi_k^{R}\!\ket{0}\bigr\rvert^{2}
\delta\bigl(k+P_{\ket{f}}-P_{\ket{0}}\bigr)
\,\delta\bigl(\epsilon +E_{\ket{f}} - E_{\ket{0}}\bigr).
\label{2.5}
\end{equation}
Here $P_{\ket{f}}$ is the momentum of the final state relative to $-p_F$.
In the absence of interactions, the matrix element in \Eq{2.5}
is finite only for a single hole excitation with momentum
$k<0$, and $A_k\propto\theta(-k)\,\delta(\epsilon-\xi_k^R)$.
Interaction results in a finite $A_k(\epsilon)$ at $\epsilon\neq \xi_k^R$.
Indeed, consider the state [see \Fig{Fig_states}(c)]
\begin{equation}
\ket{f} = \psi_{k_1}^{R}\psi_{k_2}^{L}\psi_{k_3}^{L\dagger}\ket{0},
\quad
k_1<0\,,~k_2>0\,,~k_3<0.
\label{state3}
\end{equation}
In zero order in interaction, the energy of this state is
\begin{equation}
E_{\ket{f}}-E_{\ket{0}}
=v(-k_1+k_2-k_3) - \frac{1}{2m}\left(k_1^2+k_2^2-k_3^2\right).
\label{2.6}
\end{equation}
To find the spectral edge [i.e., the largest possible
value of $\epsilon$ in \Eq{2.5} at a given $k$], we look for the
lowest possible energy of the final state $E_{\ket{f}}$ in \Eq{2.6}.
Given the momentum conservation, $k_1+k_2-k_3=k$, this limit is
reached at $k_1=k$, $k_2=k_3 = 0$ for $k<0$, and at $k_2=k$,
$k_1=k_3 = 0$ for $k>0$. In other words, the lowest-energy final
state (hence the highest possible $\epsilon$) coincides with the energy 
of a single hole, see Figs. \ref{Fig_states}(c) and \ref{Fig_states}(d).  
Consideration of final states with more than one particle-hole pair 
excited does not change this conclusion.  Accordingly, the kinematic
boundary for the hole part of the spectral function at small $k$
is given by
\begin{equation}
\overline{\xi_k} = - v|k| + k^2\!/2m.
\label{xiover}
\end{equation}
Just as it is the case for $\epsilon>0$, \Eq{xiover}
is valid for $|k|<2p_F$; finding the kinematic boundary for higher
momenta requires consideration of final states
with more excitations.

The support of the spectral function in $(p,\epsilon)$-plane is
illustrated in \Fig{Fig1}.

\subsection{Spectral function near the edges of the support}

Near the edges of support in $(p,\epsilon)$-plane,
the spectral function displays a power-law dependence on the
distance to the edge. The behavior of $A_k(\epsilon)$ near the
hole mass-shell,
\begin{equation}
A_k(\epsilon)\propto
(\overline{\xi_k}-\epsilon)^{\gamma_k^2-1}\,\theta(\overline{\xi_k}-\epsilon),
\quad -2p_F<k<0,
\label{2.20}
\end{equation}
is very similar to that in the TL model. The deviation of the exponent in
\Eq{2.20} from $(-1)$ is quadratic in the interaction and positive, just like
it is in the TL model.

Adding an extra hole to an interacting system is accompanied
by the creation of ``soft'' particle-hole pairs; we considered an example
of such three-particle final state in the previous subsection, see \Eq{state3}
and \Fig{Fig_states}(c). Excitation of multiple low-energy
particle-hole pairs leads to the orthogonality catastrophe~\cite{Anderson},
thus the correction to $(-1)$ in the exponent is positive and is proportional to
$\gamma_k^2$. The main difference compared with the TL model is
that now the extra hole interacts with the soft pairs residing near both
Fermi points, while in the TL model only pairs on the opposite branch
of the spectrum contribute to the singularity.

When the edge of the spectral support no longer coincides with the
mass-shell, which is always the case for particles ($\epsilon>0$)
and at $0<k<2p_F$ also for holes ($\epsilon<0$), the behavior
of the spectral function near the edge is different from that
described by \Eq{2.20}.  We find that the corresponding exponents are
\textit{positive} and finite even in the limit $\gamma_k\to 0$.

To be definite, let us consider the particle edge
$\epsilon\to\underline{\xi_k\!}$ at $k<0$.
The spectral function differs from zero in the vicinity of the edge 
due to three-particle final states
$\ket{f} =
\psi_{k_1}^{R\dagger}\psi_{k_2}^{L\dagger}\psi_{k_3}^{L\pdag}\ket{0}$,
see \Eq{state2} and \Fig{Fig_states}(b).
If we replace the matrix element in \Eq{2.1} by a constant, then
the dependence of $A_k(\epsilon)$ on $\epsilon-\underline{\xi_k\!}$
can be found by power counting,
\begin{eqnarray}
A_k(\epsilon)&\propto&\!\int dk_1dk_2dk_3\,\delta (k_1+k_2-k_3-k)
\nonumber\\
&& \quad\times\,\,\delta(\epsilon-\xi_{k_1}^R\!-\xi_{k_2}^L\!+\xi_{k_3}^L)
\nn\\
&\propto&(\epsilon-\underline{\xi_k\!})\,\theta(\epsilon-\underline{\xi_k\!}).
\label{2.100}
\end{eqnarray}

\Eq{2.100} corresponds to the final state $\ket{f}$ with two
particles close to the Fermi points $\pm p_F$ and a ``deep'' hole
with momentum (relative to $-p_F$) approaching $k$, see
\Fig{Fig_states}(b). Interaction of the two particles near
the Fermi level with the hole and with each other leads to a
logarithmic renormalization of the matrix element in \Eq{2.1}.
Similar to the X-ray edge singularity problem~\cite{Mahan}, the
renormalization occurs already in the first order in the interaction
strength and leads to
\begin{equation}
A_k(\epsilon)\propto
(\epsilon - \underline{\xi_k\!})^{1-\mu_k-\mu_{2p_F+k}+2\mu_{2p_F}}\,,
\quad
-2p_F<k<0.
\label{2.30}
\end{equation}
For a short range interaction and small momenta $|k|\ll p_F$ we have
$\mu_k\ll \mu_{2p_F}=2\gamma_0$, and \Eq{2.30} simplifies to
\[
A_k(\epsilon)\propto(\epsilon - \underline{\xi_k\!})^{1+2\gamma_0}.
\]
For a linear spectrum (TL model), the left-moving particle
and hole have the same velocity, destroying the core-hole effect
leading to \Eq{2.30}.

\begin{figure}[h]
\includegraphics[width=0.9\columnwidth]{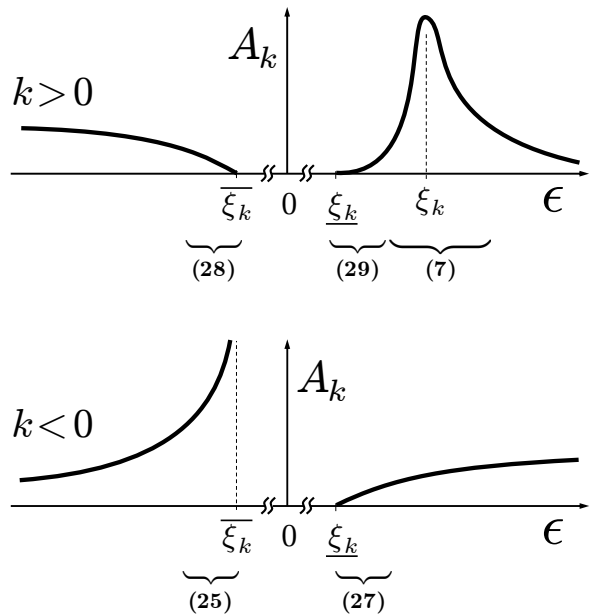}
\caption{
Dependence of the spectral function on $\epsilon$ at a fixed $k=p-p_F$
with $k>0$ (upper panel) and $k<0$ (lower panel).
The numbers in brackets refer to the equations describing
the corresponding asymptotes of $A_k(\epsilon)$.
At large $|\epsilon - \underline{\xi_k\!}|, |\epsilon - \overline{\xi_k}|\gg k^2\!/m$
(this region is not shown in the figure) the conventional Luttinger liquid behavior
$A_k(\epsilon) \propto
\left|\epsilon + vk\right|^{\gamma_0^2}\!\left|\epsilon-vk\right|^{\gamma_0^2-1}$
is restored.
}
\label{fig_sketch}
\end{figure}

The consideration of spectral function near the hole edge
$\epsilon\to\overline{\xi_k}$ at $k>0$ is very similar, yielding
 \begin{equation}
A_k(\epsilon)\propto
(\overline{\xi_k}-\epsilon)^{1-2\mu_{2p_F}-\mu_k+\mu_{2p_F-k}}\,,
\quad
0<k<2p_F.
\label{2.40}
\end{equation}
The signs of two terms in the exponent here are different from those 
in \Eq{2.30} because of the difference in the structure of the relevant 
final states. In the three-body sector, for example, the final state consists 
of one particle and two holes (rather than two particles and one hole as 
in the case of \Eq{2.30} above), see \Fig{Fig_states}(c). For a 
short-range interaction and small momenta, \Eq{2.40} simplifies to
\[
A_k(\epsilon)\propto(\overline{\xi_k}-\epsilon)^{1-2\gamma_0}.
\]

Finally, we discuss the particle edge $\epsilon\to\underline{\xi_k\!}$
at $k>0$. A new element here is that the relevant soft excitations
reside on the same branch of the spectrum. When $\epsilon$ is close
to $\underline{\xi_k\!}$, these excitations tend to occupy almost
identical right-moving states with $k\to 0$.  In the specific example
of the final state shown in \Fig{Fig_states}(a), the momenta of
two particles differ by $\sim(\varepsilon-\underline{\xi_k\!})/v$.
This results in a suppression of the matrix element in \Eq{2.1},
$\bigl\lvert\bra{f}\psi_k^{R\dagger}\!\ket{0}\bigr\rvert
\propto\varepsilon-\underline{\xi_k}\,$.
A proper modification of \Eq{2.100} then leads to
$A_k(\epsilon)\propto (\epsilon-\underline{\xi_k\!})^3$.
Accounting for the particle-hole interaction in the final state yields
\begin{equation}
A_k(\epsilon)\propto
(\epsilon - \underline{\xi_k\!}\,)^{3-2\mu_k}\,,
\quad
0<k<2p_F.
\label{2.50}
\end{equation}
At $k=2p_F$ (i.e. $p=3p_F$) the spectral edge $\underline{\xi_k\!}$
touches zero, $\underline{\xi_{2p_F}}=0$. Close to this point the
exponent can be evaluated using the conventional bosonization
technique~\cite{HaldanePRL1981}. It yields $3-4\gamma_0$ for the
exponent in agreement with \Eq{2.50} (indeed, $\mu_{2p_F}=2\gamma_0$).

The dependence of the spectral function on $\epsilon$ at a fixed $k$
is sketched in \Fig{fig_sketch}.

\subsection{Spectral function near the particle mass-shell
\label{mass-shell}
}

For a nonlinear spectrum \eq{dispersion} the particle mass-shell
$\epsilon=\xi_k^R>0$ lies \textit{above} the lower edge of
the spectral support, $\underline{\xi_k\!}= \xi_k^R - k^2\!/m$.
Far away from the mass-shell the difference between $\xi_k^R$
and $\underline{\xi_k\!}$ is not important, and at
$\epsilon-\xi_k^R\gg k^2\!/m$ the spectral function $A_k(\epsilon)$
approaches the TL form \Eq{TL}. In the vicinity of the mass-shell,
however, $A_k(\epsilon)$ undergoes a dramatic change. Indeed, since
$\epsilon=\xi_k^R$ now lies within a broader continuum, the
quasiparticle relaxation (decay) is no longer prohibited by the
conservation laws. As a result, in a parametrically wide region
(indicated by dark gray in \Fig{Fig1}) the quasiparticle peak in
$A_k(\epsilon)$ acquires a Fermi-liquid-like Lorentzian shape,
see \Eq{Lorentzian}.

\begin{figure}[h]
\includegraphics[width=0.99\columnwidth]{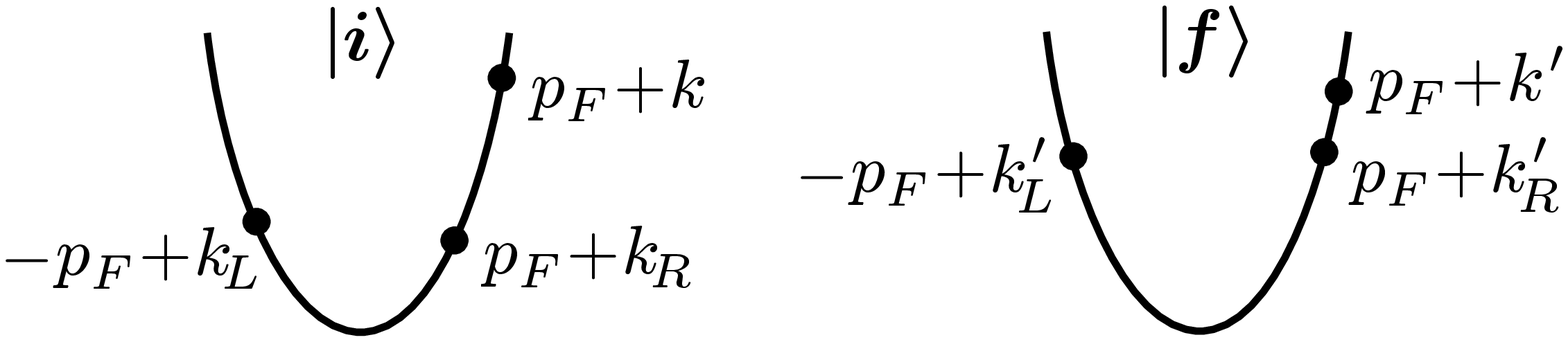}
\caption{
The initial (left panel) and the final (right panel) states of a 
three-particle scattering process that leads to a quasiparticle 
relaxation in one dimension.
}
\label{Fig_rate}
\end{figure}

It is instructive to discuss the origin of a finite relaxation rate
in a 1D Fermi system. First of all, unlike in higher dimensions,
relaxation in 1D can not occur via two-particle collisions.
Indeed, conservation of the momentum and energy allows at most
a \textit{permutation} of the momenta of two colliding particles.

Scattering processes that result in a \textit{redistribution} of the
momenta and thus potentially lead to a finite relaxation rate
must involve at least three particles. In such three-bodycollision
three particles with momenta $k$, $k_R$, and $k_L$ (relative to
$\pm\,p_F$) in the initial state $\ket{i}$ end up in a final state
$\ket{f}$ with \textit{different} momenta $k^\prime$, $k^\prime_R$,
and $k^\prime_L$, see \Fig{Fig_rate}. For a \textit{generic}
interaction~\cite{footnote} the transition $|i\rangle\to|f\rangle$ has
a nonvanishing amplitude $\mathcal A$.

In order to estimate the relaxation rate of an extra right-moving
particle with momentum $k$ due to three-body collisions,
we note that the single-particle states $k_R,k_L$
in the initial state of the transition $|i\rangle$ are below
the Fermi level, while all three single-particle states in the final
state $|f\rangle$ are above it. Applying now the Golden Rule,
we find
\begin{eqnarray}
\frac{1}{\tau_k} &\propto\!\!& \int_0^\infty\!dk' dk_R^\prime dk_L
\!\! \int_{-\infty}^0 \!dk_L^\prime dk_R \,|\mathcal{A}|^2
\label{II.C.1}
\\
&&\qquad
\times\,\delta\bigl[\left(k+k_R+k_L\right)
- \left(k'+k_R^\prime+k_L^\prime\right)\bigr]
\nn\\
&&\qquad
\times\,\delta\left[
\left(\xi_k^R+\xi_{k_R^\pprime}^R+\xi_{k_L^\pprime}^L\right)
-
\left(\xi_{k^\prime}^R+\xi_{k_R^\prime}^R+\xi_{k_L^\prime}^L\right)
\right],
\nn
\end{eqnarray}
where $\mathcal{A}$ is the three-body collision amplitude
introduced above, and the $\delta$-functions express the energy
and momentum conservation.

In writing \Eq{II.C.1} we took into account that for $k\ll p_F$
the conservation laws cannot be satisfied unless the collision
involves both the right- and the left-moving particles. Further
analysis shows that the conservation laws allow a small
$(\lesssim\! k^2\!/m)$ momentum transfer to the left-movers.
Such solution can be found by iterations. To zero order in
$k_L^\pprime-k_L^\prime$, the momentum conservation gives
$k-k'=k_R^\prime -k_R^\pprime$. The energy released in the collision
of two right-moving particles then is
$\xi_k^R+\xi_{k_{\!R}^\pprime}^R-\xi_{k^\prime}^R-\xi_{k_R^\prime}^R
\lesssim k^2\!/m$.
This energy is transferred to the left-movers,
$\xi_{k_L^\prime}^L - \xi_{k_L^\pprime}^L\!\lesssim k^2\!/m$, which
corresponds to the momentum transfer $k_L^\pprime-k_L^\prime\lesssim
k^2\!/mv \ll k$. Accordingly, the energy and momentum conservation
restrict the range of the momenta contributing to the integral in
\Eq{II.C.1} to
\begin{equation}
k', k_R^\prime,|k_R^\pprime|\lesssim k,
\quad
k_L^\pprime,|k_L^\prime|\lesssim k^2\!/mv.
\label{II.C.2}
\end{equation}

The $\delta$-functions in \Eq{II.C.1} remove the integrations over
$k_R^\prime$ and $k_L^\prime$. With the phase space constraints
\eq{II.C.2} taken into account, the remaining integrations then yield
the estimate
\begin{equation}
1/\tau_k\propto |\mathcal{A}|^2 k^4.
\label{II.C.3}
\end{equation}

For a weak generic\cite{footnote} interaction, the nonvanishing
3-particle collision amplitude $\mathcal A$ appears in the second order
in the interaction strength. More careful consideration
(see \Sec{Sec:PertTheor}) which accounts for the indistinguishability
of the two right-moving particles participating in the collision
results in the estimate
\begin{equation}
\mathcal{A}\propto V_0 (V_0 - V_k) \propto k^2,
\label{II.C.4}
\end{equation}
cf. \Eq{1dselfenergy}. Accordingly, the quasiparticle decay rate
scales with $k$ as $1/\tau_k\propto k^8$.

\section{Perturbation theory}
\label{Sec:PertTheor}

In this section we evaluate the imaginary part of the self-energy
$\im\Sigma_p(\epsilon)$ perturbatively in the interaction
strength. Specifically, we focus on the near vicinity of the
mass-shell $\epsilon\approx \xi_p$. The Fermi liquid theory
predicts a nonzero self-energy already in the second order
in the interaction strength, see \Eq{selfenergy}. We show below
that in one dimension the second-order contribution to
$\im\Sigma_p(\epsilon)$ vanishes on the mass-shell, even if the
curvature of the dispersion relation \Eq{dispersion} is taken into
account. A finite quasiparticle decay rate appears only in the
fourth order, and only if $m^{-1}\neq 0$.

We describe interacting spinless fermions by the Hamiltonian
\begin{equation}
H= \sum_{\alpha,k} \xi^\alpha_k\, \psi^{\alpha\dagger}_k
\psi^\alpha_k +
\frac{1}{2L}\sum_{q \neq 0}\left[ V_q\!\tsum_{\alpha}
\rho^\alpha_q \rho^\alpha_{-q} + 2U_q\,
\rho^R_q\rho^L_{-q} \right]  \, ,
\label{eq:1005}
\end{equation}
where $\alpha = R,L$, the dispersion relation
$\xi^\alpha_k$ is given by \Eq{dispersion}, and $\rho_q^\alpha =
\sum_{k} \psi^{\alpha\dagger}_{k-q} \psi^\alpha_k$ is the Fourier
component of the density operator. We found it convenient
to distinguish between the interbranch and the intrabranch interaction
potentials (denoted by $U_q$ and $V_q$, respectively)\cite{footnote}.
However, for the sake of brevity, we will set $U_q=V_q$ in the results.

\subsection{Second order}
\label{Sec_second}

Evaluation of the self-energy in the second order of perturbation
theory at $|\epsilon|>|\xi_p|$ parallels the corresponding
calculation in the problem with linear dispersion relation
(Tomonaga-Luttinger model). The only finite contribution comes
from the interbranch interaction $U$, and the
result reads
\begin{equation}
-\Sigma^{(2)}_p(\epsilon) =
\gamma_0^2\,(\epsilon - \xi_p)\left[\frac{1}{\pi}
\ln\frac{|\epsilon^2 - \xi_p^2|}{\epsilon_F^2} +i
\theta(|\epsilon| - |\xi_p|)\right]
\label{eq:PT1006}
\end{equation}
with $\gamma_0=\nu U_0/2$. Vanishing of the imaginary
part of the self-energy below the mass-shell is due to kinematic
constraints: the phase space available for scattering process vanishes
in the limit $\epsilon \rightarrow \xi_p$.

For $|\epsilon|<|\xi_p|$ real decay processes are allowed by conservation
laws only if the dispersion is nonlinear. For clarity, we consider
the decay of a right-moving particle ($\xi_k^R>0$).
In the presence of the spectrum nonlinearity $\im\Sigma_k$ acquires
an additional contribution in the second order in the intrabranch
interaction $V$.
This contribution comes from the scattering processes with two
particles and one hole in the final state, all three on the right
moving branch. The Golden-rule expression for $\im\Sigma_k$
reads
\begin{equation}
-\im\Sigma_k^{(2)}(\epsilon) =   \frac{\pi}{2}\!
\sum\limits_{k_1,k_2} \left| \mathcal{ A }^{(1)}_{V} \right|^2
 \delta(\epsilon -\xi_{k_1}^R - \xi_{k_2}^R + \xi_{k_3}^R)\,,
\label{eq:PT1009}
\end{equation}
where $k=p-p_F$ and $k_1+k_2-k_3 = k$ due to momentum
conservation. The Fermi statistics dictates that $k_1,k_2>0$
and $k_3<0\,$. The amplitude of the scattering process
is given by
\begin{equation}
 \mathcal{A}^{(1)}_{ V } =  V_{k-k_1} - V_{k-k_2}\, .
\label{eq:PT1012}
\end{equation}
The conservation laws are satisfied only in the limited energy
range, $\underline{\xi_k\!}\leq\epsilon\leq\xi_k^R$ with
$\underline{\xi_k\!}=\xi_k^R - k^2\!/m$, see \Eq{xiunder}. We
defer the discussion of the spectral function near the lower edge
of the spectrum $\epsilon=\underline{\xi_k\!}$ to
\Sec{Sec:Edges}, and focus here on the immediate vicinity
of the mass-shell, $|\epsilon-\xi_k^R|\ll k^2\!/2m$. In this limit
\Eq{eq:PT1009} yields
\begin{eqnarray}
-\im\Sigma_k^{(2)}(\epsilon)
= \frac{\mu_k^2}{4}\, \left(\xi_k^R -
\epsilon \right)\,\theta(\xi_k^R  - \epsilon)
\label{eq:616}
\end{eqnarray}
with $\mu_k$ defined in \Eq{mu} and $\xi_k^R$ given by \Eq{dispersion}.
The real part of the self-energy is given by
\begin{eqnarray}
-\re\Sigma_k^{(2)}(\epsilon) = \frac{\mu_k^2}{4 \pi }\left(\epsilon
- \xi_k^R
  \right) \ln\frac{\left|\epsilon - \xi_k^R
\right|}{k^2\!/m}\,.
\label{eq:618}
\end{eqnarray}

According to \Eq{eq:616}, the intrabranch interaction along with
the positive curvature of the dispersion relation ($1/m>0$) results
in a finite $\im\Sigma_k^{(2)}$ at $\epsilon<\xi_k^R$. \Eq{eq:616}
complements \Eq{eq:PT1006}, familiar from the conventional TL
theory. Note that in both cases $\im\Sigma_k^{(2)}(\epsilon)$
vanishes on the mass-shell due to the phase space constraints.

For the particle branch of spectrum ($\epsilon\approx\xi_k^R>0$), the
kinematic restrictions on the phase space are lifted in higher-order
processes.
The simplest process leading to a finite self-energy on the mass-shell
is the second-order process that results in the creation
of two particle-hole pairs in the final state.

\subsection{Fourth order}
\label{Sec:FourthOrd}

There are two kinds of the fourth order processes. The first one leads 
to a logarithmic correction to the scattering amplitude via virtual creation 
of a particle-hole pair. Similar logarithmic corrections appear in higher 
orders of perturbation theory as well. Summation of the leading logarithmic 
contributions in all orders (the corresponding procedure is described in
\Sec{Sec:LogRenorm}) results in a power-law behavior of the spectral 
function. This is very similar to the TL model, although with the exponent 
slightly modified due to the dispersion nonlinearity.

\begin{figure}[h]
\includegraphics[width= 0.95\columnwidth]{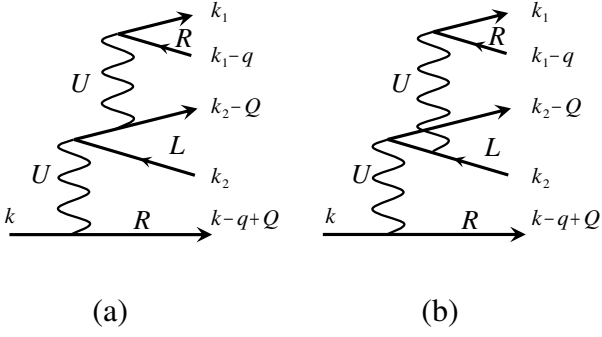}
\caption{ Second order in $U$ contributions to the amplitude
$\mathcal{ A }^{(2)}_{UU}$. (a) and (b) correspond, respectively, to
the first and second terms in \Eq{eq:PT1116}. Two more contributions 
(not shown in the figure) correspond to the replacement 
$k_1\leftrightarrow k-q+Q$ in the final states; these are the third 
and the fourth terms in \Eq{eq:PT1116}. 
}
\label{Fig:AmplRL4}
 \end{figure}

\begin{figure}[h]
\includegraphics[width=0.95\columnwidth]{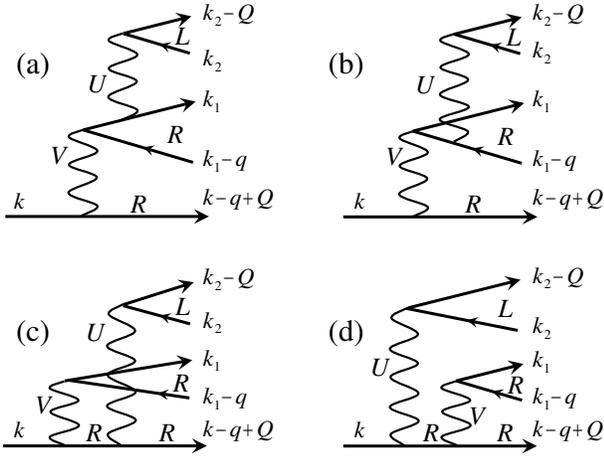}
\caption{ Second order contributions to the amplitude
$\mathcal{A}^{(2)}_{UV}$.
(a)-(d) correspond to the first terms in the numerators
of the four contributions in \Eq{eq:PT1128}. Four more
contributions (not shown in the figure) correspond to the
replacement $k_1\leftrightarrow k - q + Q$ in (a)-(d).
}
\label{Fig:AmplRR4}
 \end{figure}

A different kind of second-order processes, leading to the finite on-shell 
value of $\im\Sigma_k$,  involve creation of five quasiparticles in the final 
state: a particle on the right-moving branch, and two particle-hole pairs. 
Kinematic considerations show that the two pairs must be excited on the 
opposite branches of the spectrum in order to yield $\im\Sigma_k\neq 0$ 
on the shell. The corresponding contribution to the self-energy is then 
given by
\begin{widetext}
\begin{equation}
-\im \Sigma^{(4)}_k (\epsilon) = \frac{\pi}{2}
\sum\limits_{q,Q,k_1,k_2} \left| \mathcal{ A }^{(2)} \right|^2
\delta( \epsilon - \xi_{k-q+Q}^R + \xi_{k_1 - q}^R - \xi_{k_1}^R -
\xi_{k_2 - Q}^L + \xi_{k_2}^L )\, , 
\label{eq:PT1104}
\end{equation}
\end{widetext}
where the summation range is limited by the Pauli principle
constraints
\[
k - q + Q ,\,  k_1, \,  k_2    > 0;
\quad
k_1-q, \,  k_2 - Q  < 0,
\]
and the amplitude $\mathcal{ A }^{(2)}$ consists of two
contributions,
\begin{equation}
\mathcal{ A }^{(2)} 
= \mathcal{ A }^{(2)}_{UU} + \mathcal{A}^{(2)}_{UV}\, . 
\label{eq:PT1110}
\end{equation}
These two contributions correspond to two possible ways the desired
final state can be reached in a second-order process: The first one,
$\mathcal{ A }^{(2)}_{UU}\propto U^2,$ arises solely due to the
interbranch interaction $U$, see \Fig{Fig:AmplRL4}. The second one,
$\mathcal{ A }^{(2)}_{UV}\propto UV$, involves both the interbranch
$(U)$ and the intrabranch $(V)$ interactions, see \Fig{Fig:AmplRR4}.
The corresponding analytical expressions read
\begin{widetext}
\begin{eqnarray}
\mathcal{ A }^{(2)}_{UU} &= & \frac{ U_q U_{q - Q} }{ \xi_{k_1}^R -
\xi_{k_1 - q}^R + \xi_{k_2 - Q}^L  - \xi_{k_2 + q -Q}^L }+ \frac{
U_q U_{q - Q} }{  \xi_{k_1 - q}^R - \xi_{k_1}^R + \xi_{k_2}^L  -
\xi_{k_2 - q }^L }
\notag \\
&& - \frac{ U_{k - k_1 + Q} U_{k - k_1} }{\xi_{k-q+Q}^R - \xi_{k_1 -
q}^R + \xi_{k_2 - Q}^L  - \xi_{k_2 + k - k_1}^L }- \frac{ U_{k - k_1
+ Q} U_{k - k_1} }{   \xi_{k_1 - q}^R  - \xi_{k-q+Q}^R  +
\xi_{k_2}^L  - \xi_{k_2 - Q + k_1 - k}^L }\,. \label{eq:PT1116} \\
\nn \\
\mathcal{ A }^{(2)}_{UV} &= & \frac{ U_Q ( V_{q - Q} - V_{ k - k_1 +
Q} ) }{\xi_{k_1 }^R - \xi_{k_1 - Q}^R - \xi_{k_2}^L + \xi_{k_2-Q}^L
} +
\frac{ U_Q ( V_{q - Q} - V_{ k - k_1} ) }{\xi_{k_1 - q }^R -\xi_{k_1
- q + Q}^R + \xi_{k_2}^L - \xi_{k_2-Q}^L } +
\notag\\
&& +
 \frac{ U_Q ( V_{q} -
V_{ k - k_1} ) }{\xi_{k -q +Q }^R - \xi_{k - q}^R - \xi_{k_2}^L +
\xi_{k_2-Q}^L }+
\frac{ U_Q ( V_{q} - V_{ k - k_1 + Q } ) }{\xi_{k-q+Q}^R - \xi_{k_1
- q}^R + \xi_{k_1}^R - \xi_{k + Q}^R }\,. 
\label{eq:PT1128}
\end{eqnarray}
\end{widetext}
(A more general but rather cumbersome expression for the
amplitude free from the simplifying assumption~[\onlinecite{footnote}]
is derived in Ref.~[\onlinecite{lunde}].)

One immediately notices that $\mathcal{ A }^{(2)}_{UV}$ 
vanishes for  the momentum-independent intrabranch interaction
$V_q=V_0$. Although it is not obvious, the amplitude $\mathcal{ A
}^{(2)}_{UU}$ vanishes~\cite{Ussiskin2006} for momentum-independent
interbranch interactions $U_q=U_0$ and for a strictly parabolic
dispersion relation.

To proceed further, we assume that both inter- and intrabranch
interactions are symmetric analytic functions of the transferred
momentum; at small momenta
\begin{equation}
U_q \approx U_0 + \frac{1}{2}\,U''_0 q^2\,,
\quad
V_q \approx
V_0 + \frac{1}{2}\,V''_0 q^2\, ,
\label{eq:PT1122}
\end{equation}
and at large $q$ the potentials vanish sufficiently
fast~\cite{footnote}.  We also neglect cubic and higher-order terms in
the dispersion relation \Eq{dispersion}.

We will evaluate \Eq{eq:PT1104} for $\epsilon=\xi_k^R$.
In this limit the typical momentum of the right-moving particles in
the final states contributing to $\im \Sigma^{(4)}_k$ is of the order
of $k$ and the velocity variation is $\sim k/m$. The energy gain
due to the production of the right-moving particle-hole pairs is
therefore $\sim k(k/m)$. Conservation of energy
then yields the estimate for the typical momenta of the
left-moving particle and hole in the final state,
\[
k_2\sim Q \sim k^2\!/m v \ll k .
\]
Carrying out the summation over $k_1$ in Eq.~\eqref{eq:PT1104},
we obtain
\begin{equation}
-\im \Sigma^{(4)}_k \bigl(\xi_k^R\bigr) = \sum_{q, Q}\,\sum_{0< k_2<Q}\!
 \frac{m}{4q}
\left| \mathcal{ A }^{(2)} \right|^2\, , \label{eq:App3019}
 \end{equation}
where the summations over $q$ and $Q$ are restricted to the domain
\begin{equation}
0 < q < k,
\quad
\frac{ (k -q) \,q }{ 2 m v } < Q < \frac{ k q }{ 2 m v }.
\label{eq:App3037a}
\end{equation}
The amplitude $\mathcal{ A }^{(2)}$, see Eqs. \eq{eq:PT1110}-\eq{eq:PT1128},
simplifies considerably in the limit $k\ll mv$. Keeping only linear in 
$U''_0$ and $V''_0$ contributions, we find
\begin{equation}
\mathcal{ A }^{(2)} = U_0\left( 2 U''_0 + V''_0 \right)\, 
\frac{ q^4 - \left( 2 m v Q \right)^2}{8 m v^2 q^2} .
\label{eq:App3046}
\end{equation}
The high powers of the momenta here as well as the factor $1/m$ resulted
from delicate cancellations among various contributions to the
amplitude. Since the amplitude \eq{eq:App3046} is
independent of $k_2$, the integration over this variable in \Eq{eq:PT1104}
brings about a factor $Q/2\pi$. The remaining
integration over $q$ and $Q$ is restricted to the domain $D$
defined in \Eq{eq:App3037a},
\[
-\im \Sigma^{(4)}_k \bigl(\xi_k^R\bigr) = \int_{D} \frac{d q\, dQ}{4
\pi^2}\, \frac{mQ}{8\pi q}\, \left|\mathcal{ A }^{(2)}\right|^2
\]
The integration here is straightforward and yields
\begin{equation}
-\im \Sigma^{(4)}_k \bigl(\xi_k^R\bigr) =
\frac{ 3 U_0^2 \left(2 U''_0 + V''_0 \right)^2 \,k^8 }
{5\left(32 \pi m v^2\right)^3 }\, \, .
\label{eq:PT1140}
\end{equation}
Note that $\im \Sigma^{(4)}_k \!\bigl(\xi_k^R\bigr)$ scales with $k$
as $k^8$. The factor $k^4$ here originates in the amplitude:
according to \Eq{eq:App3046}, its typical value is $\mathcal{ A
}^{(2)}\propto k^2$. The remaining factor $k^4$ comes from the
integration over $Q$ (recall that $Q\propto k^2$). Using now
$k^2V''_0\approx 2(V_k-V_0)$, see \Eq{eq:PT1122}, and setting 
$U_k = V_k$, we arrive at \Eq{1dselfenergy} with 
$1/2\tau_k=-\im\Sigma^{(4)}_k\bigl(\xi_k^R\bigr)$ 
[here we took into account that $\im \Sigma^{(2)}_k(\xi_k)=0$].

The above derivation can be extended to $\epsilon \neq \xi_k^R$.
The self-energy varies with $\epsilon$ on the scale of the order
of $k^2\!/2m$, and it vanishes identically for
$\epsilon <\underline{\xi_k\!}$, as expected from the kinematic
considerations of \Sec{Sec:Discussion}.

According to the developed perturbation theory the spectral
function in the vicinity of the particle mass-shell takes the form
\begin{equation}
A_k(\epsilon) = \frac{1}{\pi} \,\frac{-\im
\Sigma_k(\epsilon)}{\bigl(\epsilon-\xi_k^R\bigr)^2 + \left[\im
\Sigma_k(\epsilon)\right]^2}\, ,
\label{spec_dens_perturb}
\end{equation}
where $\Sigma_k(\epsilon) = \Sigma_k^{(2)}(\epsilon)+
\Sigma_k^{(4)}(\epsilon)$ and the corresponding contributions are
given by Eqs.~\eq{eq:PT1006}, \eq{eq:616}, and \eq{eq:PT1140}.
As a result, the spectral function in the energy interval
\[
-(\tau_k\mu_k^2)^{-1}
\lesssim \epsilon-\xi_k^R
\lesssim(\tau_k\gamma_0^2)^{-1}
\]
around the mass-shell is a Lorentzian with the width
$1/2\tau_k$. Outside this interval
$\im\Sigma_k^{(2)}(\epsilon) > \im\Sigma_k^{(4)}(\epsilon)$
and the spectral function is described by a power law
$A_k(\epsilon)\sim |\epsilon-\xi_k^R|^{-1}$, with the exponent
equal to $(-1)$. The perturbation theory developed so far does not
take into account the logarithmic renormalization of the
scattering amplitudes. In \Sec{Sec:LogRenorm} we show that
summation of the leading logarithmic corrections leads to the
interaction-dependent correction to the exponent.
The height of the Lorentzian peak \Eq{spec_dens_perturb}, is also
renormalized, see \Eq{Lorentzian} above.

\section{Leading logarithmic corrections }
\label{Sec:LogRenorm}

It is easy to see that excitation of virtual particle-hole pairs
leads to the logarithmically divergent contributions in perturbation theory,
in a very much the same way as in the TL model. The real part of the self-energy
acquires such logarithmic corrections already in the second order in
interaction, see \Eq{eq:PT1006}. The logarithmic terms in the imaginary part
of the self-energy appear in the fourth order. [These contributions
vanish on the mass-shell, and thus do not affect the validity of \Eq{eq:PT1140}].

In this section we develop a procedure for the summation of the
leading logarithmic corrections to the Green function $G_k(\epsilon)$
in the presence of nonlinear terms in the dispersion relation \Eq{dispersion}.
The nonlinearity is not important as long as $|\epsilon-\xi_k^R|\gg k^2/2m$.
Close to the mass-shell, however, the behavior of the spectral function
deviates significantly from that in the conventional Luttinger liquid. 
For a positive curvature [i.e. for $1/m>0$ in \Eq{dispersion}], deviations 
are the strongest for the particle ($\epsilon>0$) excitations with $k>0$; 
we have already seen that the on-shell excitations acquire a finite 
lifetime $\tau_k$. On the other hand, the behavior of the hole branch 
($\epsilon<0$) is qualitatively similar to that of a Luttinger liquid. 
We will concentrate here on the properties of the particle branch 
at $|\epsilon-\xi_k|\ll k^2/m$, deferring the discussion of the hole region 
of the spectrum till the end of this Section.

\subsection{\!\!\!
Vicinity of the particle mass-shell:
$\bm\epsilon\bm\to\bm\xi_{\bm k}^{\bm R},~\bm k\bm>\bm 0$
}
\label{Sec:particlemassshell}

In order to account simultaneously for both the logarithmic
renormalization and the finite quasiparticle lifetime, we notice
that the relevant energy scales form a well-defined hierarchy
\begin{equation}
\frac{1}{2\tau_k} \ll \frac{k^2}{2m} \ll \xi_k^R\ll mv^2.
\label{scalesepration}
\end{equation}
The logarithmic corrections originate from almost the entire
energy band, i.e. from the states with energies in the range
$(|\epsilon-\xi_k^R|, mv^2)$. On the other hand, the finite lifetime $1/\tau_k$
originates in the decay of a particle into particle and hole states within
much narrower strip of energies of the width of the order of $\xi_k^R$.

 \begin{figure}[h]
 \includegraphics[width=0.8\columnwidth]{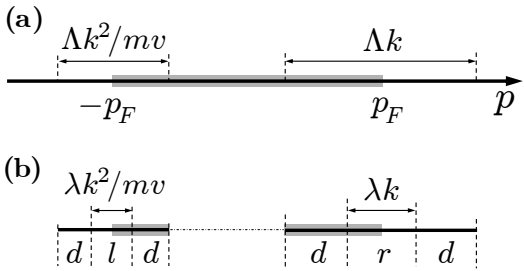}
 \caption{
(a) Subbands in the momentum space remaining after performing the
 RG transformation described in step (i) in the text.
(b) Subbands of the effective model described by the
Hamiltonian~\eq{splitH}. The state with $k>0$ under
consideration belongs to $d$-subband.
}
\label{Fig:Subbands}
 \end{figure}

Our strategy for evaluation of the Green
function will be as follows:

(i) First we will take into account virtual transitions to states with
relatively high energies, corresponding to the momenta $p$ in the range
$\Lambda k\lesssim |p-p_F|\lesssim p_F$
for the right-movers and $\Lambda k^2\!/mv\lesssim |p+p_F|\lesssim p_F$
for the left-movers (here $\Lambda\gg 1$). This step is similar to the 
standard renormalization group (RG) calculation for the Tomonaga-Luttinger 
model~\cite{Solyom}. As a result of ``integrating out'' the high-energy 
states, $G_k(\epsilon)$ acquires a multiplicative factor. In addition, some 
irrelevant terms are generated in the Hamiltonian; these terms do not affect 
the logarithmic renormalization of $G_k(\epsilon)$, but contribute to the 
decay rate $1/\tau_k$ which is evaluated on the latter stage. The multiplicative 
RG terminates when the energy bands have been reduced to strips of the 
right-moving states with $|p-p_F|\lesssim\Lambda k$, and the left-movers 
with $|p+p_F|\lesssim\Lambda k^2\!/mv$, see \Fig{Fig:Subbands}(a).

(ii) In the next step, we separate the remaining states in three groups.
The first two (subbands $r$ and $l$) correspond to two segments,
$|p-p_F|\lesssim\lambda k$ and $|p+p_F|\lesssim\lambda k^2\!/mv$
on the right- and left-moving branches, respectively;
hereinafter $\lambda\ll 1$. The rest of the states
form subbands $d$, see \Fig{Fig:Subbands}(b).
The state of interest $k$ belongs to the right-moving branch of $d$-subband.
This state acquires a finite lifetime $\tau_k$, which
manifests itself in the shifting of the pole in $G_k(\epsilon)$
by $i/2\tau_k$ off the real axis. Importantly, the dominant contribution to
the finite lifetime comes from the decay of the state of interest into other
states within $d$-band.
Although the subbands $r$ and $l$ make a negligible contribution
to the decay rate $1/\tau_k$, the density fluctuations in these subbands
induce a slowly varying [on the timescale of the order of $(k^2\!/2m)^{-1}$]
fluctuating field that affects the dynamics of the high-energy
($\varepsilon\sim\xi_k^R$) $d$-particle.
Our formalism accounts for both these fluctuations and for
the finite decay rate $1/2\tau_k$, and leads to \Eq{result}.

Now we sketch the implementation of the two steps described above. 
(Some technical details are relegated to 
Appendixes \ref{Sec:AppVerCor} and \ref{Sec:AppDerivIntegr}.)

Step (i) follows, with a small modification, the conventional RG
procedure employed in the theory of Tomonaga-Luttinger
model~\cite{Solyom}. We introduce a reduced space of
single-particle states with $|p-p_F|<k_R$ for
the right-movers and $|p+p_F|<k_L$ for the left-movers and
consider the Green function, $G_k(\epsilon; k_L,k_R)$, defined in
the reduced band. Following the idea of the multiplicative
RG~\cite{Solyom}, we consider the transformation of the Green
function associated with the reduction of the bandwidth
($k_L\to k_L^\prime$, $k_R\to k_R^\prime$), and cast it in the
form
\begin{equation}
G_k(\epsilon; k_L^\prime,k_R^\prime)
=z\left(\frac{k_L^\prime}{k_L},\frac{k_R^\prime}{k_R}\right)
G_k(\epsilon; k_L,k_R)
\label{rg-transf}
\end{equation}
Next, we use the results of the \Sec{Sec_second}
to evaluate the Green function \eq{rg-transf} perturbatively in the
second order in the interaction potential. This leads to an approximate
expression for $z$,
\begin{equation}
z\left(\frac{k_L^\prime}{k_L},\frac{k_R^\prime}{k_R}\right)
=1-2\gamma_0^2\cdot\ln\frac{k_L^\prime}{k_L}-0\cdot\ln\frac{k_R^\prime}{k_R},
\label{approx-z}
\end{equation}
where $\Lambda k <k^\prime_{L,R}<p_F$.  The last term on the
right-hand side here expresses the fact that the
renormalization of $G_k(\epsilon)$ for a right-mover results from
its interaction with the left-movers; this is why the function $z$
depends only on a single argument $k_L^\prime/k_L$.
\Eq{approx-z} is valid up to the second order in
$\gamma_0$, and also assumes that $\gamma_0^2\ln(k_L^\prime/k_L)\ll
1$. In order to find $z(k_L^\prime/k_L)$, we supplement \Eq{approx-z}
by the requirement that the scaling function has a multiplicative
property~\cite{Solyom}. That leads to
\begin{equation}
z\left(\frac{k_L^\prime}{k_L},\frac{k_R^\prime}{k_R}\right)
=\left(\frac{k_L^\prime}{k_L}\right)^{-2\gamma_0^2}\!,
\quad
\Lambda k<k_{R,L},k_{R,L}^\prime \!<p_F .
\label{z}
\end{equation}

In order to keep the state of interst $k$ under consideration
it must lie within the reduced band of right-movers. Accordingly,
the scaling of the right band must stop at $k_R^\prime\approx\Lambda k$.
However, the bandwidth of the left-movers can be reduced even further,
all the way down to $k_L^\prime \approx \Lambda k^2\!/mv$.
The latter scale represents the width of the band to which the states
involved in real transitions associated with the formation of $1/\tau_k$
belong, see \Sec{Sec:FourthOrd}. (Note that according to
\Eq{eq:PT1006}, the scaling exponent in the interval
$\Lambda k^2\!/mv <k_L^\prime <\Lambda k$ is $\gamma_0^2$ rather
than $2\gamma_0^2$).

Using now Eqs.~\eq{rg-transf} and \eq{z}, we find
\begin{equation}
G_k(\epsilon)= \left(\frac{\Lambda
kv}{\epsilon_F}\right)^{2\gamma_0^2} \left(\frac{\Lambda
k^2/m}{\Lambda kv} \right)^{\gamma_0^2}
G_k\left(\epsilon;\frac{\Lambda k^2}{mv}, \Lambda k\right)
\label{reduced}
\end{equation}
(we used $mv\sim p_F$ here). Note that the nonlinearity of
the single-particle spectrum does not affect the form of
Eqs.~\eq{approx-z} and \eq{reduced} as long as $\Lambda\gg 1$.

However, because of the nonlinearity, reduction of the bandwidth
($k_F\to \Lambda k^2/p_F$) generates an additional interaction term
\begin{equation}
{\hat V}_\Lambda=\sum M \,
\psi_{k_1}^{R\dagger}\psi_{k-q-Q}^{R\dagger} \psi_{k_2 -
Q}^{L\dagger}
\psi_{k_2}^{L}\psi_{k}^{R}\psi_{k_1-q}^{R} \, .
\label{a15}
\end{equation}
The matrix element $M$ here accounts for the contributions to the
total transition amplitude $\mathcal{ A }^{(2)}$ of the virtual
states outside the reduced band of the left-movers. These states
contribute to $\mathcal{ A }^{(2)}_{UU}$, but not to $\mathcal{ A
}^{(2)}_{UV}$. Therefore one finds $M\approx (1/2)\mathcal{ A
}^{(2)}_{UU}$ (here $1/2$ is a combinatorial factor). Since all the
momenta in \Eq{a15} belong to the reduced band, we have
\begin{equation}
{M} = U_0  U''_0 \,
\frac{q^4 - \left( 2 m v Q\right)^2}{8 m v^2 q^2} .
\label{a15A}
\end{equation}
as it is clear from the analysis leading to Eq.~\eqref{eq:App3046}.
Using $q\sim k$ and $Q \sim k^2/p_F$, we estimate
$M\sim k^2\!/m v^2 $, i.e., $M$ vanishes
for a linear spectrum. The generated interaction \Eq{a15} is to be added
to the Hamiltonian \Eq{eq:1005},
where all the operators now act within the band of reduced width.
Along with the interaction terms already present
in  \Eq{eq:1005}, ${\hat V}_\Lambda$ contributes to the inelastic scattering
processes giving rise to a finite quasiparticle relaxation rate $1/\tau_k$.

We now proceed to step (ii) of the program outlined above,
i.e., evaluation of the Green function in the reduced band.
To this end, we separate the states in the reduced band
in two narrow subbands:
the $l$-band with $|k^\ast|\lesssim \lambda k^2/p_F$ around the left Fermi point,
and the $r$-band with $|k^\ast|\lesssim \lambda k$ around the right
Fermi point. The remaining states form subband $d$, see
\Fig{Fig:Subbands}(b). Parameter $\lambda$ here satisfies
the conditions
\begin{equation}
\gamma_0^2\ln\frac{\Lambda}{\lambda}\ll 1,
\quad
\lambda\ll 1\,.
\label{lambda}
\end{equation}
(Note that at $\gamma_0\ll 1$ the two conditions \eq{lambda} can be 
satisfied simultaneously.)
The states within the subbands $r$ and $l$ produce additional logarithmic
renormalization of the Green function. The first condition in
\Eq{lambda} allows us to disregard such corrections originating in
$d$-subband. The second condition in \Eq{lambda} ensures
that accounting for the inelastic scattering within $d$-subband reproduces
the result of \Sec{Sec:PertTheor} for the relaxation rate $1/\tau_k$.

We write the Hamiltonian $H+{\hat V}_\Lambda$ acting in the reduced
subspace $r,l,d$ (see Fig~\ref{Fig:Subbands}) as
\begin{equation}
H+{\hat V}_\Lambda= H_d+H_{rl}+H_{d-rl}.
\label{splitH}
\end{equation}
The first and second terms in the right-hand side here account for the
states in ``high-energy'' subband $d$ and the low-energy subbands
$r$ and $l$, 
\[
H_d = P_d(H+{\hat V}_\Lambda)P_d\,,
\quad H_{rl}=P_{rl}(H+{\hat V}_\Lambda)P_{rl}\,,
\]
where $P_d$ and $P_{rl}$ are projectors onto the corresponding states.
The third term in \Eq{splitH} describes the interactions,
\[
H_{d-rl}=P_d(H+{\hat V}_\Lambda)P_{rl}
+P_{rl}(H+{\hat V}_\Lambda)P_d .
\]
Using Hamiltonian Eq.~(\ref{splitH}), we evaluate the Green
function of a particle in the subband $d$.

We start the evaluation of $G_k\left(\epsilon;\frac{\Lambda
k^2}{mv},  \Lambda k\right)$ with accounting for the interaction terms in
$H_d$. Evaluation of the self-energy follows the perturbative
analysis of \Sec{Sec:PertTheor}. The only difference is that the 
second-order contribution to $\im\Sigma_k$, see \Eq{eq:616}, is absent 
provided that $|\epsilon-\xi_k^R|<\lambda k^2\!/mv$. 
Indeed, a finite $\im\Sigma_k^{(2)}$ requires excitation of a particle-hole
pair in the $l$-subband. Such processes are absent in $H_d$ but are 
included in $H_{d-rl}$ discussed below. In addition, the second-order
contribution to the real part of the self-energy, 
$\re\Sigma_k\sim \gamma_0^2(\epsilon-\xi_k^R)\ln(\Lambda/\lambda)$, 
can be also neglected as it is small compared with $|\epsilon-\xi_k^R|$. 
Unlike in the second order, the low-energy states ($l$ and $r$ subbands) 
do not play any special role in the fourth order calculation, see  
\Sec{Sec:PertTheor}.
Hence, $H_d$ alone is sufficient in order to reproduce the relaxation
rate $1/2\tau_k$ (note that the term ${\hat V}_\Lambda$ is
important for this calculation). Thus, neglecting $H_{d-rl}$, we
find for the retarded Green function
\begin{equation}
G^d_k\left(\epsilon;\frac{\Lambda k^2}{mv},
  \Lambda k\right)=\frac{1}{\epsilon-\xi_k^R+i/2\tau_k}\, .
\label{gf-d}
\end{equation}

Next, we account for the effects of $H_{d-rl}$ interaction on the
Green function of a particle in the $d$-subband. For this calculation
${\hat V}_\Lambda$ is irrelevant. Moreover, the interaction between
$r$ and $l$ subbands does not contribute to the Green function
in the leading logarithmic approximation (this interaction leads merely
to the higher-order corrections to the exponents).
We thus neglect both these terms and write $H_{rl}$ and $H_{d-rl}$ as
\begin{eqnarray}
H_{rl} &=& \sum_{\alpha=r,l}\sum_k \xi^\alpha_k\,
\psi^{\alpha\dagger}_k \psi^\alpha_k
\label{drl}
\\
H_{d-rl} &=& (V_0-V_k)\frac{1}{L}\!\sum_{|q| < \lambda k}  \rho^d_q
\rho^r_{-q}
\label{drl111}
\\
&&\quad+\, (U_0-U_{2p_F+k})\frac{1}{L}\!
\sum_{|q| < \lambda k^2\!/mv}
\rho^d_q \rho^l_{-q}  \, ,
\nn
\end{eqnarray}
where the form of the interaction potentials $V_0-V_k$ and $U_0-U_{2p_F+k}$ 
originates in the reduction of the general interaction to the density-density form. 
Because of the strict limitation on the wavelengths of the density fluctuations,  
the subbands $r$ and $l$ do not contribute to $1/\tau_k$. Therefore the spectrum 
within these subbands can be linearized. Since both $r$ and $l$ subbands 
contain the Fermi level, the excitations within these subbands can be described
using the conventional bosonization technique. Upon introducing the bosonic 
fields $\varphi_q^{r,l}$ such that $\rho_q^{r,l}= iq\varphi_q^{r,l}/ 2 \pi $, 
we represent $H_{rl}$ and $H_{d-rl}$ in the form
\begin{widetext}
\begin{eqnarray}
H_{rl}&=&\frac{1}{4\pi L}\sum_{|q| < \lambda k}vq^2
\left|\varphi_q^r\right|^2  + \frac{1}{4\pi L}\sum_{|q| < \lambda
k^2\!/mv}vq^2  \left|\varphi_q^l\right|^2
\label{hrl}\\
H_{d-rl}&=& - i\frac{V_0-V_k}{2\pi L}\sum_{|q| < \lambda k}q
\rho_q^d \varphi_{-q}^r  - i \frac{U_0-U_{2p_F+k}}{2\pi
L}\sum_{|q| < \lambda k^2\!/mv} q \rho_q^d \varphi_{-q}^l\, .
\label{hdrl}
\end{eqnarray}
\end{widetext}
In the second order of perturbation theory the interactions in $H_d$ 
can be neglected. The self-energy of $d$-particle is then given by the diagram 
shown in Fig. 9(a). Its evaluation reproduces the logarithmic divergence at 
$\epsilon\to\xi_k^R$, see \Eq{eq:618}.

\begin{figure}[h]
 \includegraphics[width=0.65\columnwidth ]{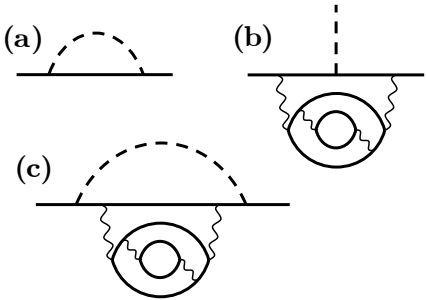}
 \caption{
(a) lowest-order logarithmic correction to the $d$-particle
propagator (solid line); the dashed line stands for $r(l)$ bosons.
(b) example of renormalization of the $d$-particle-boson vortex;
wavy lines represent interaction within the $d$-subband.
(c) one of the contributions to the decay rate of a $d$-particle.
}
\label{Fig:Dressing}
 \end{figure}

To go beyond the lowest order, we need to take into account
simultaneously both the interactions present in $H_d$
as well as the interactions of $d$-particles with the bosons.
Since we are interested in the limit $\epsilon-\xi_k^R\to 0$,
it is sufficient to consider only the most divergent in this limit 
contributions.

Clearly, accounting for both types of interactions produces two
kinds of contributions. First, interactions within the $d$-subband
renormalize the particle-boson interaction vertex. Vertex corrections 
such as that in \Fig{Fig:Dressing}(b) can be shown to be
small (see Appendix~\ref{Sec:AppVerCor}), and we neglect them 
hereinafter. Second, the interactions ``dress'' the bare $d$-particle 
Green function, see \Fig{Fig:Dressing}(c), replacing it by that given 
in \Eq{gf-d}. Obviously, the effect of the dressing is to cut the logarithmic
divergencies off at $|\epsilon-\xi_k^R|\sim 1/\tau_k$.

The task of evaluating contributions such as that shown in
\Fig{Fig:Dressing}(c) is greatly simplified by the fact that
particle-boson interaction is associated with rather small
$(\lesssim\lambda k)$ transferred momentum. This allows us to
linearize the $d$-particle spectrum about $\xi_k^R$,
\[
\xi_{k'}^d = \xi_{k+k'}^R \approx \xi_k^R + v_d k',
\quad
v_d = v+k/m.
\]

It is convenient to write the $d$-particle Green function in
the real space-time representation,
\[
G_k\left(\epsilon; \frac{\Lambda k^2}{mv}\,,\Lambda k\right)
=\int\!dx\!\int\!dt\,
e^{i\epsilon t}G_k^d(\mathbf{x}),
\quad
\mathbf{x}\equiv (x,t).
\]
In the absence of particle-boson
interaction $G_k^d$ satisfies the equation
\[
 \left( i \frac{\partial}{\partial t} + i  v_d\,\frac{\partial}{\partial x}
  - \xi_k^R +
  \frac{i}{2\tau_k}  \right)\,
  G_k^d(\mathbf{x})=\delta( \mathbf{x})\, .
\]
With vertex corrections neglected, the effect of the bosonic fields
is merely to induce a slowly varying in space and time potential in
which the $d$-particle moves. This fluctuating potential can be written as
\begin{equation}
\phi(x,t) = \frac{V_0-V_k}{2 \pi}\, \partial_x \varphi^r +
\frac{U_0-U_{2p_F+k}}{2 \pi}\,\partial_x \varphi^l\,.
\label{HdPotential}
\end{equation}
The retarded Green function of $d$-particle in the presence of
the fluctuating potential $G_k^d(\mathbf{x}|\phi)$ satisfies the equation
\begin{equation}
  \left( i \frac{\partial}{\partial t} + i  v_d\,\frac{\partial}{\partial x}
  - \xi_k^R +
  \frac{i}{2\tau_k} + \phi(\mathbf{x}) \right)\,
  G_k^d(\mathbf{x}|\phi)=\delta( \mathbf{x})\,,
\label{Greenequation}
\end{equation}
and $G_k^d(\mathbf{x})$ is obtained by averaging $G_k^d(\mathbf{x}|\phi)$
over the gaussian fluctuations of the field $\phi$,
\[
G_k^d(\mathbf{x})=\langle G_k^d(\mathbf{x}|\phi)\rangle_\phi\,.
\]
Here $\langle\ldots\rangle_\phi$ denotes the time-ordered averaging over 
the slowly varying field $\phi$, see \Eq{HdPotential}. Note that $G_k^d$ 
in \Eq{Greenequation} describes propagation of a single $d$-particle in 
an empty band, hence the corresponding retarded Green function coincides 
with a time ordered one. 

Carrying out the calculations, we find
\begin{equation}
G_k^d\left(\mathbf{x}|\phi\right) =
- i \theta(t)\, \delta( x - v_d t )\,e^{ -i \xi_k^R t  -  t / 2
\tau_k}\, e^{ i [\theta(\mathbf{x})-\theta(\mathbf{0})] }
\label{Greensolution1}
\end{equation}
with
\begin{equation}
\theta(\mathbf{x}) =  \int \frac{ d \omega }{ 2 \pi }
 \sum_{q}  \frac{ i \phi(q,\omega) }{\omega - v_d\, q }
\,\, e^{ i q x - i \omega t}\, ;
\label{Phase}
\end{equation}
the summation over $q$ here is restricted to $|q|<\lambda k$ for
\protect{$r$-$d$} interactions and to $|q|<\lambda k^2\!/mv$ for 
\protect{$l$-$d$} interactions. Note that because $\theta$ enters 
\Eq{Greensolution1} only in the combination $\theta(v_dt,t)-\theta(0,0)$, 
the pole at $\omega=v_d q$ in \Eq{Phase} does not show up in 
\Eq{Greensolution1}.

With the dynamics of the bosonic fields described by a quadratic 
Hamiltonian \Eq{hrl}, the averaging in Eqs. \eq{Greensolution1}
and \eq{Phase} is straightforward.  It yields
\begin{equation}
 G_k^d(\mathbf{x}) =
- i \theta(t)\, \delta( x - v_d t )\,
e^{-i \xi_k^R t - \frac{t}{2\tau_k}-K(\mathbf{x} )}\, ,
\label{Greenaverage}
\end{equation}
with
\begin{eqnarray}
K(\mathbf{x}) &=&  \frac{1}{2}\left\langle \left(\theta(\mathbf{x})
- \theta(\mathbf{0})\right)^2 \right\rangle_\phi
\label{PhaseFluct}
\\
 &= &  \int \frac{ d \omega }{ 2 \pi }
 \sum_{q}  \frac{\langle\phi(\mathbf{q})\phi(\mathbf{-q})\rangle_\varphi } {(\omega - v_d\, q)^2 }\,\,
 \left(1  - \cos(q x - \omega t) \right) \,.
\nn
\end{eqnarray}
Here
\begin{eqnarray}
 &&\langle\phi(\mathbf{q})\phi(\mathbf{-q})\rangle_\phi
 \nonumber \\
  && = i (U_0-U_{2p_F+k})^2\, \Pi^{l} (\mathbf{q}) + i (V_0
  - V_k)^2 \, \Pi^{r} (\mathbf{q}),
\label{phi_phi}
\end{eqnarray}
and 
\[\Pi^{r,l} (\mathbf{q})= - i \langle
\rho^{r,l}\rho^{r,l}\rangle_\phi = (q/2\pi)\bigl(\pm \,\omega - v q + i
0 \sign q\bigr)^{-1}
\] 
are density-density correlation functions for $r$
and $l$ bosons. Substitution of \Eq{phi_phi} into \Eq{PhaseFluct}
and then into \Eq{Greenaverage} yields
\begin{widetext}
\begin{equation}
G_k^d  (\mathbf{x})=
- i \theta(t) \delta( x - v_d t ) \, e^{ -i \xi_k t  -  t / 2
\tau_k}
\exp\left\{
-\frac{\mu_{2p_F+k}^2}{4} \ln\left[1 + i \frac{\lambda k^2}{mv} (x +
v t ) \right] - \frac{\mu^2_k}{4} \ln\left[1 - i \lambda k (x - v t )
\right]
 \right\}\,.
\label{Greensolution}
\end{equation}
The Green function in the reduced band is given by the
Fourier transform of this result (see Appendix~\ref{Sec:AppDerivIntegr}
for the details of the calculation),
\begin{equation}
G_k\left(\epsilon;\frac{\Lambda k^2}{mv}, \Lambda k\right) =
\frac{1}{\epsilon-\xi_k^R+{i/ 2\tau_k} }
  \left[ \frac{\mu_{2p_F+k}^2}{4\gamma_k^2} \,
  \left(\frac{-\epsilon+\xi_k^R-{i/
  2\tau_k}}{\lambda k^2/m} \right)^{\gamma_k^2} +
\frac{\mu_k^2}{4\gamma_k^2} \,
\left(
\frac{\epsilon-\xi_k^R+i/2\tau_k}{\lambda k^2\!/m}
\right)^{\gamma_k^2} \right].
\label{FT}
\end{equation}
Using now \Eq{reduced} and taking into account conditions
\Eq{lambda}, we finally arrive at the expression  for the spectral 
function [the abbreviated versions of it were given in \Eq{result} 
and \Eq{Lorentzian}]
\begin{equation}
A_k(\epsilon) =
\frac{1}{4\pi}\,
\left(\frac{\xi_k^R}{\epsilon_F^2}\right)^{\!\!\gamma_0^2}
\left(\frac{k^2}{m}\right)^{\gamma_0^2-\gamma_k^2}\, \im
\left\lbrace \frac{\mu_{2p_F+k}^2}{\gamma_k^2}\, \left(
\frac{-1}{\epsilon-\xi_k^R+i/2\tau_k} \right)^{1-\gamma_k^2}\!
- \,\, \frac{\mu_{k}^2}{\gamma_k^2} \, \left(
\frac{1}{\epsilon-\xi_k^R+i/2\tau_k} \right)^{1-\gamma_k^2}
\right\rbrace.
\label{fullresult}
\end{equation}
\end{widetext}

The validity of the result \eq{fullresult} is restricted to the vicinity
of the particle mass-shell, $|\epsilon- \xi_k^R|\lesssim \lambda k^2\!/2m$.
It is clear however that well above the mass-shell, at $\epsilon-\xi_k^R \gg
k^2\!/2m$, the dispersion nonlinearity has no effect and the
spectral function crosses over to the conventional TL expression
\Eq{TL}. On the other hand, $\epsilon=\underline{\xi_k\!}=\xi_k^R-k^2/m$
below the mass-shell represents the kinematic edge of the spectrum,
see \Sec{Sec:Discussion}.
Evaluation of the spectral function in the immediate vicinity
of the edge requires a special consideration, and is discussed in
\Sec{Sec:Edges} below.

\subsection{
Vicinity of the hole mass-shell:
$\bm\epsilon\bm\to{\bm\xi_{\bm k}^{\bm R}}$, $\bm k\bm<\bm 0$
}
\label{Sec:holemassshell}

The behavior of the spectral function near the hole mass-shell
($\epsilon\to\xi_k^R<0$) is much simpler because the decay of a hole
is prohibited by the energy and momentum conservation laws.
The spectral function can be evaluated by following closely the
route outlined in \Sec{Sec:particlemassshell} apart from two
important modifications:
First, one needs to replace $1/\tau_k\to +i0$ in
\Eq{Greenequation} and its solution. Second,
unlike $d$-particle, $d$-hole has a velocity $v_d$ which is smaller
than $v$; the difference is due to the positive curvature of the dispersion
relation.
This affects the analytical properties of
\Eq{Greensolution}. Indeed, at $x=v_dt$ the
factors $x+vt$ and $x-vt$ have the same sign for $v_d<v$. This in turn
results in $A_k(\epsilon)=0$ at $\epsilon>\xi_k^R$, which
agrees with the kinematic constraints on the hole part
of the spectral function, see \Sec{Sec:Discussion}.
Below the mass-shell, at $0<\xi_k^R-\epsilon\ll k^2\!/2m$, we find
\begin{equation}
G_k(\epsilon)=
\left(\frac{|\xi_k^R|}{\epsilon_F^2}\right)^{\!\!\gamma_0^2}\,
\frac{(k^2/m)^{\gamma_0^2-\gamma_k^2}} {\epsilon-\xi_k^R+{i0} }\,
  \left(\frac{\epsilon-\xi_k^R+{i0}}{k^2/m}
  \right)^{\gamma_k^2}\,.
\label{FThole}
\end{equation}
Imaginary part of this expression is given in \Eq{Ahole}.

\section{Edge singularities}
\label{Sec:Edges}

In the previous section we found that the spectral function
diverges at the hole mass-shell, which coincides with the edge of
the spectrum $\epsilon = \overline\xi_k$ at $k<0$. Here we
consider the behavior of the spectral function in the vicinity of
the remaining kinematic boundaries. As discussed in
\Sec{Sec:Discussion}, in all cases the spectral function exhibits
a power-law suppression at the edge. The suppression originates in
the phase space constraints which lead to vanishing of the spectral
function linearly with the distance to the edge. Interactions modify
the exponent via a mechanism analogous to the X-ray edge singularity
in metals\cite{Mahan}: the transition amplitudes in Eqs. \eq{2.1} and
\eq{2.5} acquire a power-law dependence on the distance to the edge.
In this Section we develop a technique to account for this dependence.

The edge of the spectrum corresponds to final states of the
transition in which all of the momentum and energy are carried by
a single hole, see \Sec{Sec:Discussion}. At energies close to, but
not precisely at, the kinematic edge, the final states may
contain, in addition to this ``deep hole'', an arbitrary number of
low-energy particle-hole pairs near the two Fermi points.

When the distance in energy to the corresponding edge is small
compared to $k^2\!/m$, the hole can be treated as distinguishable
from the rest of the particles in the system\cite{Pustilnik2006,MP_CSM}.
Formally, this amounts to projecting out all states except those in the
narrow stripes of momenta $r,l$ near the Fermi points, and a strip $d$
deep below the Fermi level. (Note the difference between the narrow
$d$-subband defined in this Section and the wide one in \Sec{Sec:LogRenorm}).
The $d$-subband hosts a single hole, and the interaction of this hole with
the rest of the system results in the excitation of the particle-hole pairs in
the subbands $r$ and $l$.

We consider first the behavior of the spectral function near the hole
edge $\epsilon\approx {\overline\xi_k}$ at $k>0$,
and then proceed with the consideration of the particle edge
$\epsilon\approx \underline{\xi_k\!}$.

\subsection{
$\bm\epsilon\bm\to\overline{\bm\xi_{\bm k}}\,$, $\bm k\bm>\bm 0$
}

\begin{figure}[h]
\includegraphics[width=0.5\columnwidth]{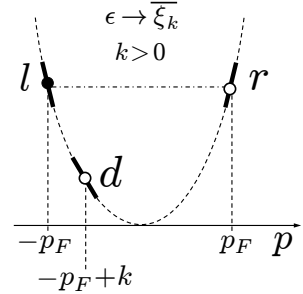}
\caption{
Solid lines indicate the states included in the effective Hamiltonian
\eq{5.0}-\eq{5.3}. The $d$-subband hosts a single hole. The interaction
of the hole with $r$ and $l$-subbands leads to the excitation of
low-energy particle-hole pairs. Production of multiple particle-hole pairs
results in a power-law dependence of the spectral function in a close
analogy with X-ray edge singularity in metals.
}
\label{Fig_hole_states}
\end{figure}

In this case the deep hole in the final state of the transition is
located at momentum close to $k$ (relative to $-p_F$) on the
left-moving branch of the spectrum, see \Fig{Fig_hole_states}. In
order to describe the final-state interaction, we project the
Hamiltonian \eq{eq:1005} onto narrow strips of momenta shown in
\Fig{Fig_hole_states}, $H_{\rm eff} = {\cal P}H{\cal P}$, where
$\cal P$ projects onto states within $r,l$, and $d$-subbands,
while the remaining states are regarded as either occupied or
empty. In order to extract the dependence of the spectral function
on the distance to the edge with the leading logarithmic accuracy,
it is sufficient to carry out the projection to zero order
in interactions for the Hamiltonian and to the lowest nonvanishing
(first) order for the observable. The omitted higher order contributions
would contribute beyond the leading logarithmic accuracy.

We introduce slowly varying in space fields
\begin{equation}
\psi^{r,l}(x) = \sum_{|k|<k_0} \!\frac{e^{ikx}}{\sqrt{L}} \,\psi^{R,L}_k,
\quad
\psi^d(x) = \sum_{|k'\!|<k_0}\! \frac{e^{ik'x}}{\sqrt{L}} \,\psi^{L}_{k+k'}
\label{x-space}
\end{equation}
where $k_0\ll k$ is the high-momentum cutoff, and write the
effective Hamiltonian in the coordinate representation as
\begin{equation} H_{\rm eff} = H_{rl} + H_{d} + H_{\rm int}.
\label{5.0}
\end{equation}
Here
\begin{equation}
H_{rl} =\int\!dx\left[ \psi^{r\dagger}(-iv\partial_x)\,\psi^r
+ \psi^{l\dagger}(iv\partial_x)\psi^l\right]
\label{5.1}
\end{equation}
and
\begin{equation}
H_d = \int\!dx \,\psi^{d\dagger}(\overline{\xi_k} + iv_d\,\partial_x)\psi^d
\label{5.2}
\end{equation}
describe $r,l$, and $d$ subbands, respectively.
In writing Eqs. \eq{5.1} and \eq{5.2} we linearized the spectrum within
the respective subbands, so that
\begin{equation}
\xi^L_{k+k'} = \overline{\xi_k} -v_d k',
\quad
v_d = v -k/m.
\label{5.05}
\end{equation}
The third term in the right-hand side of \Eq{5.0} describes the interaction
between the subbands, 
\begin{eqnarray} H_{\rm int} 
&= & - \int\!dx\,\overline\rho^{\,d}\! 
\left[(V_0-V_k)\rho^{\,l} + (V_0-V_{2p_F-k})\rho^r\right]
\nn\\
&&\quad + ~(V_0-V_{2p_F})\!\int\!dx\,\rho^r(x)\rho^{\,l}(x),
\label{5.3}
\end{eqnarray}
where $\rho^{r,l}(x) = \psi^{r,l\,\dagger}(x)\psi^{r,l}(x)$ are particle densities
in $r,l$-subbands, and $\overline\rho^{d}(x) = \psi^{d}(x)\psi^{d\dagger}(x)$ is
density of holes in $d$-subband. In writing \Eq{5.3} we have set $V_q=U_q$.

Our goal is to evaluate the hole contribution to the spectral function
\begin{equation}
A_k(\epsilon) 
= \re\frac{1}{\pi}\!\int^0_{-\infty} \!dt\, e^{-i\epsilon t}
\bigl\langle\psi^{R\dagger}_k(t) \,\psi^{R\pdag}_k\!\!(0)\bigr\rangle.
\label{5.4}
\end{equation}
The  projection should now be applied to the operator
$\psi^{R}_k$ in \Eq{5.4}. In this case, however, the lowest order
is insufficient, as the state with momentum $k$ on the right-moving
branch of the spectrum lies outside the subbands $r,l,d$ of the
effective Hamiltonian. Instead, the relevant contribution is generated
in the first order in the interaction strength.

The higher-order contributions can be found with the help of a unitary
(Schrief{}fer-Wolf{}f) transformation that decouples states
within the subbands $r,l$, and $d$ from the rest of the system.
Consider the following term in the original Hamiltonian \eq{eq:1005},
\begin{eqnarray}
\delta H &=& (V_k\!-\!V_{2p_F})\frac{1}{L}\!
\sum_{|k_i|<k_0}\!\!\delta_{k_1+k_2,k_3}
\label{5.01}\\
&&\qquad\times\,\,
(1-{\cal P})\,\psi^{R\dagger}_k\psi^{R\pdag}_{k_1}
\psi^{L\dagger}_{k_2}\psi^{L\pdag}_{k+k_3} \,{\cal P}
\,\,+\,\text{H.c.}
\nn
\end{eqnarray}
Schrief{}fer-Wolf{}f transformation $H\to e^{i\hat S}H e^{-i\hat S}$
eliminates the off-diagonal contributions such as \Eq{5.01}. To the
lowest (first) order in the interaction strength, the generator of such
transformation reads
\begin{eqnarray*}
\hat S &=& \frac{V_k\!-\!V_{2p_F}}{iL(\xi^R_k - \xi^L_k)}\!
\sum_{|k_i|<k_0}\!\!\delta_{k_1+k_2,k_3}
\\
&&\qquad\times\,\,
(1-{\cal P})\,\psi^{R\dagger}_k\psi^{R\pdag}_{k_1}
\psi^{L\dagger}_{k_2}\psi^{L\pdag}_{k+k_3} \,{\cal P}
\,\,+\,\text{H.c.},
\end{eqnarray*}
so that $[i\hat S,H_0] = -\delta H$, where $H_0$ is the noninteracting
part of \Eq{eq:1005}.

As far as the effective Hamiltonian \eq{5.0} is concerned, the
transformation leads merely to the correction to $H_{\rm int}$,
\[
\delta H_{\rm int} = \frac{1}{2}\bigl[i\hat S,\delta H\bigr],
\]
which amounts to negligible second-order corrections to the
coupling constants in \Eq{5.3}.

The same Schrief{}fer-Wolf{}f transformation applied to the
operator $\psi^{R}_k$ yields
\begin{eqnarray}
\psi^R_k &\to& \bigl[i\hat S,\psi^R_k\bigr]
\label{5.5}\\
&=& - \frac{V_k - V_{2p_F}}{\xi^R_k - \xi^L_k} \frac{1}{L}
\sum_{|k_i|<k_0}\!
\psi^{L\dagger}_{k_1}\psi^{R\pdag}_{k_2}\psi^{L\pdag}_{k+k_3}
\delta_{k_1\!,k_2+k_3}.
\nn
\end{eqnarray}
This contribution, as well as those generated in higher orders,
conserves separately the numbers of right- and left-movers.
With this constraint, the first-order contribution \eq{5.5} is a product of
the least possible number of operators acting in the subbands $r,l,d$,
hence, it is the most relevant one as far as the behavior of the spectral function
near the edge is concerned.

Using now Eqs. \eq{x-space}, \eq{5.4}, and
\eq{5.5}, we find
\begin{equation}
A_k(\epsilon) \propto \re\!\int\!dx\!\int^0_{-\infty} \!dt\, e^{-i\epsilon t}\!
\left\langle\Psi^{\dagger}(x,t)\Psi^{\pdag}\!\!(0,0)\right\rangle
\label{5.6}
\end{equation}
with
\begin{equation}
\Psi(x) = \psi^{l\dagger}(x)\, \psi^r(x)\, \psi^d(x).
\label{5.7}
\end{equation}

To proceed further, we bosonize fermions in $r$ and $l$-subbands according to
\begin{equation}
\psi^{r,l} = \sqrt{k_0}\, e^{\pm \,i \varphi^{r,l}},
\quad
[\varphi^{r,l}\!(x),\varphi^{r,l}\!(y)] = \pm\, i\pi\sign(x-y),
\label{5.100}
\end{equation}
where the upper/lower sign corresponds to $r/l$, and $k_0$ is high-momentum
cutoff for bosonic modes,
\[
\left\langle
\varphi^{r,l}(x)\varphi^{r,l}(0) - \left[\varphi^{r,l}(0)\right]^2
\right\rangle
= -\ln\bigl(1\mp 2\pi i k_0 x\bigr).
\]
In the bosonic representation Eqs. \eq{5.1} and \eq{5.3} read
\begin{equation}
H_{rl} = \frac{v}{4\pi}\!\sum_{\alpha=r,l}\!\int\!dx\,(\partial_x\varphi^\alpha)^2
\label{5.8}
\end{equation}
\begin{eqnarray}
H_{int} &=&
- \!\int\!\frac{dx}{2\pi}\,\overline\rho^d\!
\left[(V_0-V_k)\,\partial_x\varphi^l \!+ (V_0-V_{2p_F-k}\!)\,\partial_x\varphi^r\right]
\nn\\
&&\quad+\,\,(V_0-V_{2p_F\!})\!
\int\!\frac{dx}{(2\pi)^2}\,(\partial_x\varphi^r)(\partial_x\varphi^l)\,,
\label{5.9}
\end{eqnarray}
and \Eq{5.7} becomes
\begin{equation}
\Psi(x)= k_0\, e^{i\,[\varphi^r(x) + \varphi^l(x)]}\,\psi^d(x).
\label{5.10}
\end{equation}

The correlation function \eq{5.6} with $\Psi$ given by \Eq{5.10} and with
the dynamics governed by the effective Hamiltonian \eq{5.2}, \eq{5.8}, and
\eq{5.9} can be evaluated exactly. This can be done by following the steps
outlined in \Sec{Sec:LogRenorm}. Equivalently, one can diagonalize the
effective Hamiltonian by an appropriate unitary
transformation\cite{Pustilnik2006,MP_CSM}. The generator of such
transformation reads
\begin{eqnarray}
\widehat W &=& \frac{1}{2}\!\int\!dx\,\overline\rho_d(x)
\bigl[\mu_{2p_F\!-k\,}\varphi^r(x) - \mu_{k\,}\varphi^l(x)\bigr]
\label{5.11}\\
&&~~ +\, \frac{\mu_{2p_F}}{8\pi}\!\int\!dx\,
\bigl[\varphi^r \partial_x\varphi^l-\varphi^l \partial_x\varphi^r\bigr]
\nn
\end{eqnarray}
with $\mu_k$ introduced in \Eq{mu} above. In writing \Eq{5.11} we
omitted all but the first-order in interactions contributions
(recall that second and higher order corrections have been
neglected in the derivation of the effective Hamiltonian). The
transformation decouples the subbands from each other to linear
order in interactions,
\begin{equation}
e^{i\widehat W}H_{\rm eff\,} e^{-i\widehat W} = H_{rl} + H_d, \label{5.12}
\end{equation}
and also modifies the operator $\Psi$, see \Eq{5.10},
\begin{equation}
e^{i\widehat W}\Psi \,e^{-i\widehat W}\! \propto \psi^d\,
e^{i\left[1+ \frac{1}{2}(\mu_{2p_F-k}-\mu_{2p_F})\right]\varphi^r}
e^{i\left[1- \frac{1}{2}(\mu_{k}+\mu_{2p_F})\right]\varphi^l}
\label{5.13}
\end{equation}

Since the transformed Hamiltonian \eq{5.12} is quadratic,
evaluation of the correlation function \Eq{5.6} is straightforward.
Keeping only linear in $\mu_p$ terms in the exponents, we find
\begin{eqnarray}
\left\langle\Psi^{\dagger}(x,t)\Psi^{\pdag}\!\!(0,0)\right\rangle
\!\!&\propto&\,
\delta(x+v_dt)\, e^{i\overline{\xi_k}t}
\label{5.18}\\
&\times&\!\!k_0\bigl[1+2\pi i k_0 (vt-x)\bigr]^{-1-\mu_{2p_F-k}+\mu_{2p_F}}
\nn\\
&\times&\!\!k_0\bigl[1+2\pi i k_0 (vt+x)\bigr]^{-1+\mu_k +\mu_{2p_F}}.
\nn
\end{eqnarray}
Substitution of \Eq{5.18} into \Eq{5.6} then yields the spectral function
near the edge,
\begin{equation}
A_{k}(\epsilon) \propto
\left(\overline{\xi_k} - \epsilon  \right)^{1-\mu_{k}
+ \mu_{2 p_F -k} - 2\mu_{2 p_F}}
\theta \!\left(\overline{\xi_k} -\epsilon \right),
\label{5.19}
\end{equation}
see \Eq{2.40}. This result is valid when $\epsilon$ is close to
the edge, $\overline{\xi_k}  - \epsilon\ll k^2\!/m$, and for
$0<k<2p_F$ (note that for $k>p_F$ the above derivation should be
modified as in this case $d$-subband in the corresponding
effective Hamiltonian belongs to the right-moving branch of the
spectrum).

The interaction-induced corrections to $+1$ in the exponent in \Eq{5.19} 
(which comes from the phase-space constraints) allow for a simple 
interpretation based on the analogy with X-ray edge 
singularity\cite{Mahan} in metals: $-\mu_k$ originates in the attractive
interaction of the deep hole on the left-moving branch of the spectrum 
with a soft left-moving particle; $\mu_{2p_F-k}$ is due to the repulsion 
between the deep hole and a soft right-moving hole; finally, $-2\mu_{2p_F}$ 
is due to the attraction between a soft left-moving particle and a soft 
right-moving hole (note that both particles are soft, hence the factor 
of 2).

For small $k$, the exponent in \Eq{5.19} simplifies to
\[
1-\mu_{k} + \mu_{2 p_F -k} - 2\mu_{2 p_F}
\to 1-\mu_k - \mu_{2p_F}
\to 1-2\gamma_0,
\]
where we used $\mu_{2p_F} = 2\gamma_0$ and and the fact that for a generic
interaction $\mu_k\to 0$ when $k\to 0$.

\subsection{
$\bm\epsilon\bm\to\underline{\bm\xi_{\bm k}\!}\,$, $\bm k\bm<\bm 0$
}

\begin{figure}[h]
\includegraphics[width=0.5\columnwidth]{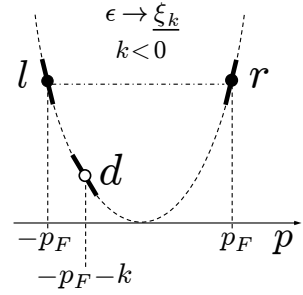}
\caption{
Solid lines indicate the subbands $r,l$, and $d$ in the effective Hamiltonian
for the evaluation of the particle contribution to the spectral function
at $\epsilon\to\underline{\xi_k\!}$, $k<0$.
}
\label{Fig_part1_states}
\end{figure}

The consideration in this case is very similar to that in \Sec{Sec:Edges}A.
The effective Hamiltonian accounting for the final-state interaction again consists
of subbands $r$ and $l$ near the two Fermi points, and $d$-subband (now
centered at momentum $-k>0$) deep below the Fermi level on the left branch
of the spectrum, see \Fig{Fig_part1_states}.
The projection, carried out to the lowest order in interactions, yields
Eqs. \eq{5.0}-\eq{5.3} with the replacement
$\overline{\xi_k}\to-\underline{\xi_k\!}$ and $v_d\to v+k/m$ in \Eq{5.2},
and $k\to-k$ in \Eq{5.3}.

The particle contribution to the spectral function is given by
\begin{equation}
A_k(\epsilon) = \re\frac{1}{\pi}\!\int_0^{\infty} \!dt\, e^{i\epsilon t}
\bigl\langle\psi^{R\pdag}_k\!\!(t)\,\psi^{R\dagger}_k(0)\bigr\rangle.
\label{5.20}
\end{equation}
Similar to above, application of the Schrief{}fer-Wolf{}f transformation to
the operator $\psi^{R\dagger}_k$ yields the relevant contribution in the first
order in the interaction strength,
\begin{equation}
\psi^{R\dagger}_k \propto  \frac{V_k - V_{2p_F+k}}{\xi^R_k +
\xi^L_{-k}} \frac{1}{L}\! \sum_{|k_i|<k_0}\!
\psi^{R\dagger}_{k_1}\psi^{L\pdag}_{k_2}\psi^{L\pdag}_{-k+k_3}
\,\delta_{k_1+k_2,k_3}. \label{5.21}
\end{equation}
When written in the coordinate representation, Eqs. \eq{5.20} and \eq{5.21} give
\begin{equation}
A_k(\epsilon) \propto\re\!\int\!dx\!\int_0^{\infty} \!dt\, e^{i\epsilon t}
\bigl\langle\Psi^{\pdag}\!\!(x,t)\Psi^{\dagger}(0,0)\bigr\rangle
\label{5.22}
\end{equation}
with
\begin{equation}
\Psi^\dagger(x) = \psi^{r\dagger}(x)\,\psi^{l\dagger}(x)\, \psi^{d\pdag}\!\!(x).
\label{5.23}
\end{equation}

The spectral function \eq{5.22} is evaluated by bosonizing the
Hamiltonian and diagonalizing it by a unitary transformation, just
as it is done in \Sec{Sec:Edges}A. This procedure yields
\begin{equation}
A_k(\epsilon)\propto
(\epsilon - \underline{\xi_k\!}\,)^{1-\mu_k-\mu_{2p_F+k}+2\mu_{2p_F}}
\theta \!\left(\epsilon-\underline{\xi_k\!}\,\right).
\label{5.25}
\end{equation}
This result is valid at $\epsilon-\underline{\xi_k\!}\ll k^2\!/m$,
and for all $k$ in the range $-2p_F<k<0$. Note that $\mu_{2p_F}$ and
$\mu_{2p_F+k}$ enter the exponent in \Eq{5.25} with opposite signs
compared to those in \Eq{5.19}. This is because the interaction of the
deep hole with a particle on the $r$-branch is \textit{attractive},
whereas the corresponding correction to the exponent in \Eq{5.19}
originates in the \textit{repulsion} between the deep hole and another
hole in $r$-subband.

As expected, the exponent in \Eq{5.25} is invariant upon the
momentum inversion $k\leftrightarrow 2p_F+k$.
At very small $|k|$, the exponent in \Eq{5.25} simplifies to
\[
1-\mu_k-\mu_{2p_F+k}+2\mu_{2p_F}
\to
1+\mu_{2p_F} = 1+2\gamma_0.
\]

\subsection{
$\bm\epsilon\bm\to\underline{\bm\xi_{\bm k}\!}\,$, $\bm k\bm>\bm 0$
}

\begin{figure}[h]
\includegraphics[width=0.5\columnwidth]{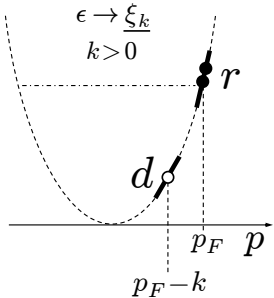}
\caption{
States included in the effective Hamiltonian \eq{5.50}-\eq{5.53}
for the evaluation of the spectral function at $\epsilon\to\underline{\xi_k\!}$, $k>0$.
}
\label{Fig_part2_states}
\end{figure}

In this limit the final state of the transition involves creation
of a hole with the momentum $-k$ on the right-moving branch of the
spectrum, see \Fig{Fig_part2_states} and discussion in
\Sec{Sec:Discussion}. In order to evaluate the spectral function
in the leading logarithmic approximation, it is sufficient to
consider the effective Hamiltonian consisting of $d$-subband
hosting the hole, and $r$-subband to allow for the creation of
low-energy particle-hole pairs near the right Fermi point, see
\Fig{Fig_part2_states}.  (Inclusion of left-movers would merely
add a second-order in interaction correction to the exponent.)
Carrying out the projection and linearizing the spectrum within
the subbands, we find
\begin{equation}
H_{\rm ef{}f} = H_{r} + H_{d} + H_{\rm int}
\label{5.50}
\end{equation}
with
\begin{eqnarray}
H_{r} &=& \int\!dx\,\psi^{r\dagger}(-iv\partial_x)\,\psi^r ,
\label{5.51}\\
H_d &=& \int\!dx \,\psi^{d\dagger}(-\underline{\xi_k\!} - iv_d\,\partial_x\!)\,\psi^d
\label{5.52}
\end{eqnarray}
(here $v_d=v-k/m$), and \begin{equation} H_{\rm int} = -
\,(V_0-V_k) \!\int \!dx\,\overline\rho^d(x) \rho^r(x).
\label{5.53}
\end{equation}

After bosonizing Eqs. \eq{5.51} and \eq{5.53}, the effective Hamiltonian
can be diagonalized by a unitary transformation with generator
\begin{equation}
\widehat W = -\frac{\mu_k}{2}\int dx\,\overline{\rho}^d(x)\,\varphi^r(x).
\label{5.54}
\end{equation}
The transformation yields, to first order in the interaction
strength, \begin{equation} e^{i\widehat W}H_{\rm ef{}f\,}
e^{-i\widehat W} = H_{r} + H_d, \label{5.55}
\end{equation}
where $H_r = (v/4\pi)\!\int\!dx\,(\partial_x\varphi^r)^2$ and $H_d$
is given by \Eq{5.52}.

Since the right-moving state with momentum $k\gg k_0$ lies outside the
domain of the effective Hamitlonian \eq{5.50}, application of the
lowest-order projection to the operator $\psi^{R\dagger}_k$, see
\Eq{5.20}, is insufficient. Similar to above, the relevant
contribution emerges in the first order in the interaction
strength and reads
\begin{equation}
\psi^{R\dagger}_k \propto
\frac{1}{L}\!\sum_{|k_i|<k_0}\!\!
\frac{V_{k-k_1} - V_{k-k_2}}{\xi^R_k + \xi^R_{-k}}\,
\psi^{R\dagger}_{k_1}\psi^{R\dagger}_{k_2}\psi^{R\pdag}_{-k+k_3}
\delta_{k_1+k_2,k_3}.
\label{5.31}
\end{equation}
Since $k_0\ll k$, we can expand here
\[
V_{k-k_1} - V_{k-k_2}\approx
(k_2-k_1)\,\frac{dV_k}{dk}\,.
\]
Passing over to the coordinate representation, we find that the
spectral function is given by \Eq{5.22} with
\begin{equation}
\Psi^\dagger(x) = \psi^{r\dagger}(x)
(-i\partial_x)\psi^{r\dagger}(x)\,\psi^d(x)\,.
\label{5.32}
\end{equation}
Bosonizing \Eq{5.32} according to \Eq{5.100} and taking proper care
of the point splitting (see Appendix~\ref{point_splitting} for the details),
we find
\begin{equation}
\Psi^\dagger(x) = 2\pi k_0^2\,e^{-2i\varphi^r(x)}\psi^d(x).
\label{5.33}
\end{equation}
The unitary transformation \eq{5.54}-\eq{5.55} that diagonalizes the
effective Hamiltonian also modifies the operator $\Psi^\dagger$:
\begin{equation}
e^{i\widehat W}\Psi^\dagger(x)\,e^{-i\widehat W}
\propto\,e^{-i(2-\mu_k/2)\,\varphi^r(x)}\psi^d(x).
\end{equation}
With the dynamics governed by the quadratic Hamiltonian \eq{5.55},
it is straightforward to calculate
\begin{eqnarray}
\bigl\langle\Psi^{\pdag}\!(x,t)\Psi^{\dagger}(0,0)\bigr\rangle
&\propto& \delta(x-v_d t)\,e^{-i\underline{\xi_k\!}\,t}
\label{5.34}\\
&&\times\,k_0^4\bigl[1+2\pi i k_0 (vt-x)\bigr]^{-4+2\mu_k}.
\nn
\end{eqnarray}
Substitution of \Eq{5.34} into \Eq{5.22} then yields
\begin{equation}
A_k(\epsilon)
\propto
\left(\epsilon -\underline{\xi_k\!}\,\right)^{3-2\mu_{k}}
\theta\!\left( \epsilon -\underline{\xi_k\!}\,\right)
\label{5.35}
\end{equation}
for the spectral function at $\epsilon-\underline{\xi_k\!}\ll
k^2\!/m$. The fact that the interaction-induced correction to the exponent
is $-2\mu_k$ is due to the presence of two soft particles interacting with 
the deep hole,  contributing $-\mu_k$ each.

\section{Calogero-Sutherland model}
\label{Sec:Calogero}

In this section we consider a solvable model of interacting 1D
fermions, the Calogero-Sutherland (CS) model. The corresponding
Hamiltonian in the first-quantized form reads~\cite{Sutherland}
 \begin{equation}
 H = -\sum_i\frac{1}{2m}\frac{\partial^2}{\partial x_i^2}\, +
\sum_{i<j} V(x_i-x_j), 
\label{CS1} 
\end{equation} 
where $V(x)$ is a periodic version of the inverse-square interaction potential,
\begin{equation} 
V(x) = \frac{\lambda(\lambda - 1)/m}{(L/\pi)^2\sin^2(\pi x/L)}\,. 
\label{CS2} 
\end{equation}

Correlation functions of the CS model exhibit rather unusual behavior.
For example, the dynamic structure factor $S(q,\omega)$ differs from
zero in a finite interval of
frequencies~\cite{ZH_SLP,CSM_hole,Drag2003,MP_CSM}, just as
it is for free fermions. On the contrary, for a generic interaction
the structure factor has a high-frequency ``tail''
$S\propto q^4/\omega^2$, which emerges already in the second
order of the perturbation theory in the interaction
strength~\cite{Drag2003,Pustilnik2006}. However, close examination
of the corresponding perturbative formula for the
structure factor (see Eq.~(18) of Ref.~[\onlinecite{Drag2003}])
indeed shows that it yields zero for $V_k \propto|k|$.

Similarly, substitution of $U_k=V_k \propto |k|$ into Eqs.
\eq{eq:PT1104}-\eq{eq:PT1128} yields
$1/2\tau_k=-\im\Sigma_k^{(4)}(\xi_k) \equiv 0$.
Just as it is the case with the absence of the high-frequency tail in
$S(q,\omega)$, the apparent vanishing of the relaxation rate in perturbation
theory suggests that $1/\tau_k=0$ is the exact relation for the CS model.
Indeed, this agrees with the exact results for the Green
function~\cite{ZH_SLP} obtained for specific values of
$\lambda$.

Vanishing of $1/\tau_k$ for particles is a peculiar property of
the CS model related to its integrability\cite{footnote}.
Therefore, in the context of this work, we
concentrate on the \textit{hole} contribution to the spectral
function; we believe that the CS model results for the hole
$(\epsilon<0)$ region of the spectrum are generic.

Single-particle correlation functions for the CS model have been
studied extensively~\cite{ZH_SLP,CSM_hole}
in the context of the exclusion statistics~\cite{Polychronakos}.
Such interpretation is possible because the inverse-square potential is
impenetrable. Requiring the many-body wave function to obey a
certain symmetry with respect to the permutations of the
particles' coordinates amounts to merely choosing a rule according
to which the wave function is assigned an overall phase. For CS
model, the phase depends on the ordering of particles, but not on
their coordinates. Operators which do not permute particles (e.g.,
the local density operator) do not affect the phase factor.
Accordingly, the statistics of bare particles is immaterial as
far as the evaluation of, say, the dynamic structure factor
$S(q,\omega)$ is concerned\cite{MP_CSM}.

However, the situation with single-particle correlation functions is
more subtle. The anyon creation and annihilation operators constructed
and studied in Ref.~[\onlinecite{CSM_hole,ZH_SLP}] describe fermions
only for odd integer values of $\lambda$. Rather than attempting to
derive fermionic Green function for general values of $\lambda$, below
we will use the results of Refs.~[\onlinecite{CSM_hole}] to evaluate
$A_k(\epsilon)$ for $\epsilon<0$ and $\lambda =$ odd integer, and then
employ an analytical continuation to extend the result to arbitrary
values of $\lambda$.

The excitations of CS model can be described in terms of
quasiparticles and quasiholes. Quasiholes are characterized by
fractional inertial mass $\bar m = m/\lambda$ and velocities $v_i$
in the range $|v_i|<v= p_F/\bar m$, where $p_F = \pi n$ is the
Fermi momentum ($n$ is particle concentration). On the contrary,
quasiparticles have velocities $|v_i|>v$ and their mass coincides
with the bare mass $m$ that enters \Eq{CS1}.

Consider now a state obtained by a removal of a single particle
(with mass $m$) from the ground state $\ket{0}$. For integer
$\lambda$, this change of mass can be accommodated by creation of
exactly $\lambda$ quasiholes. It turns out that this simplest
possibility is, in fact, exhaustive~\cite{CSM_hole}.

\begin{figure}[h]
\includegraphics[width=0.75\columnwidth]{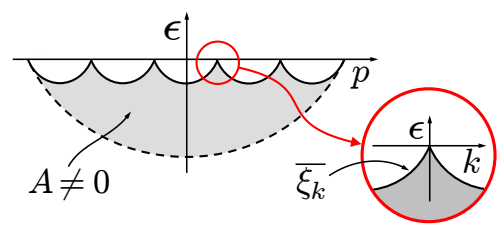}
\caption{Support of the spectral function in $(p,\epsilon<0)$
half-plane. The upper boundary of the region where $A\neq 0$
comprises of $\lambda$ identical parabolic segments ($\lambda=5$
in the figure). We concentrate on the low-energy sector
$k=p-p_F\to 0$ (see the inset).
\label{fig_CSM}}
\end{figure}

The region of support of $A_p(\epsilon)$ in $(p,\epsilon)-$plane
can now be deduced from the energy and momentum conservation.
Indeed, using the well-known expressions~\cite{Sutherland} for
the energy and momentum of a state with $\lambda$ quasiholes with
velocities $v_i$, one can write
\begin{eqnarray}
\epsilon + \sum_{i=1}^\lambda \frac{\bar m}{2} \,(v^2 - v_i^2) = 0,
\quad
p - \sum_{i=1}^\lambda \bar m v_i = 0. \label{CS4}
\end{eqnarray}
For odd integer $\lambda$, these equations have a solution for $\{v_i\}$
provided that $\epsilon$ and $p$ lie within the shaded region in
\Fig{fig_CSM}. The upper boundary of this region,  the solid
line $\epsilon = \overline{\xi_p}$, comprises of $\lambda$
identical parabolic segments,
\begin{equation}
\overline{\xi_p} =
\frac{1}{2\bar m}\bigl[(p-2l p_F)^2-p_F^2\bigr],
\quad
|p_{\,}-2l p_F|<p_F
\label{CS5}
\end{equation}
with integer $l$, $|l|\leq (\lambda-1)/2$.
The support of $A_p(\epsilon)$ is also bounded
from below by the dashed line $\epsilon  = p^2\!/2m-\epsilon_F$
with $\epsilon_F = mv^2\!/2$; this boundary would be absent for
generic values of $\lambda$.

We concentrate here on the low-energy sector with $p\approx p_F$,
$|\epsilon|\ll\epsilon_F$ (see \Fig{fig_CSM}) and ``shift'' the
momentum according to $k=p-p_F$. The dependence of $A_k(\epsilon)$
on $\epsilon$ has a threshold,
\begin{equation}
A_k(\epsilon)\propto\theta(\overline{\xi_k} - \epsilon),
\quad
\overline{\xi_k} = - v|k| + k^2\!/2\bar m, 
\label{CS6}
\end{equation}
and our goal here is to find the behavior of $A$
at a fixed $k$ when $\epsilon$ approaches the threshold. This can
be done by writing the spectral function in the form of a multiple
integral,
\begin{eqnarray}
&&
A_k(\epsilon) \propto m\! 
\int_{-v}^{v} \prod\limits_{i=1}^\lambda dv_i
\,F\bigl(\{v_i\}\bigr)
\label{CS7}\\
&&
\qquad \times \,\delta\bigl[k + p_F - \bar m\!\tsum  v_i\bigr]
\,\delta\bigl[\epsilon + (\bar m/2)\!\tsum (v^2 - v_i^2)\bigr].
\nn
\end{eqnarray}
The $\delta-$functions here reflect the conservation of
momentum and energy, see \Eq{CS4}, and the form-factor
$F\propto\bigl|\bra{\{v_i\}}\psi_k\ket{0}\bigr|^2$
was found in Ref.~[\onlinecite{CSM_hole}],
\begin{equation}
F\bigl(\{\bar v_i\}\bigr)
 = \prod_{i_1<i_2} |v_{i_1}-v_{i_2}|^{2/\lambda} \prod_i
\,\bigl(v^2-v_i^2\bigl)^{-\frac{\lambda -1}{\lambda}} .
\label{CS8}
\end{equation}
\Eq{CS7} is nothing but the Lehmann representation of the spectral 
function, with the final state of the transition $\ket{\{v_i\}}$ parametrized 
by the velocities of $\lambda$ quasiholes.

For $|k|\ll p_F$ and $\overline{\xi_k} -\epsilon\ll \epsilon_F$,
the final states $\{v_i\}$ contributing to $A$ have $n_\lambda =
(\lambda - 1)/2$ quasiholes with velocities $v_i\approx - v$ and
$n_\lambda + 1$ quasiholes with velocities $v_i\approx +v$.
It is therefore convenient to introduce new variables
\[
x_i = \frac{v-v_i}{v}\,, \quad y_i = \frac{v+v_i}{v}\,.
\]
After expansion of the form-factor to the lowest nonvanishing
order in $x_i,y_i\ll 1$, \Eq{CS7} takes the form
\begin{eqnarray} 
A_k(\epsilon)&\propto&\frac{1}{\epsilon_F} \prod_{i,j}
\!\int_0^\infty \!\!\!dx_i \!\int_0^\infty \!\!\!dy_j
\,f_{n_\lambda+1}\bigl(\{x_i\}\bigr)
\,f_{n_\lambda}\bigl(\{y_j\}\bigr) 
\nn
\\
&&\qquad\times~\delta\!\left[\frac{\epsilon
+ vk}{2\bar m v^2} + X_1 - \frac{1}{4}(X_2 + Y_2)\right]
\label{CS9}
\\
&&\qquad\times~\delta\!\left[\frac{\epsilon - vk}{2\bar m v^2} +Y_1 -
\frac{1}{4}(X_2 + Y_2)\right].
\nn
\end{eqnarray}
Here
\[
X_n = \sum_{i=1}^{n_\lambda +1} x_i^n,
\quad
Y_n = \sum_{i=1}^{n_\lambda} y_i^n,
\]
and $f_N\bigl(\{z_i\}\bigr)$ is a function of $N$ arguments
$z_1,\ldots,z_N$ given by
\begin{equation}
f_N\bigl(\{z_i\}\bigr)
= \prod_{i_1<i_2}|z_{i_1}-z_{i_2}|^{2/\lambda}
\prod_i\,z_i^{-
\frac{\lambda -1}{\lambda}}\,;
\label{CS10}
\end{equation}
this is a homogeneous function of degree
\[
c_N = \frac{1}{4\lambda}\Bigl\lbrace\bigl[2(N-n_\lambda)-1\bigr]^2
- \lambda^2\Bigr\rbrace.
\]

As it is easy to check,
\[
X_2\sim X_1^2\ll X_1\ll 1, \quad Y_2\sim Y_1^2\ll Y_1\ll 1.
\]
Moreover, when $\epsilon$ is relatively far from the threshold, at
$\overline{\xi_k} -\epsilon\gg k^2\!/2\bar m$, $X_1$ and $Y_1$ are
of the same order, $X_1\sim Y_1$. In this limit $X_2$ and $Y_2$ in
the arguments of the $\delta$-functions in \Eq{CS9} can be safely
neglected, after which the integration reduces to a power counting
which yields
\begin{equation}
A_k(\epsilon) \propto 
\left|\epsilon + vk\right|^{\gamma_0^2} \left|\epsilon
-vk\right|^{\gamma_0^2 - 1},
\quad \overline{\xi_k}
-\epsilon\gg{k^2}\!/{\bar m}
\label{CS11}
\end{equation}
with
\begin{equation}
\gamma_0^2 = \frac{(\lambda - 1)^2}{4\lambda\,}\,. 
\label{CS12}
\end{equation}
Eqs. \eq{CS11} and \eq{CS12} reproduce the standard Luttinger
liquid result.  At $|vk -\epsilon|\ll v|k|$ \Eq{CS11} agrees with
\Eq{TL} above. Since the exponent $\gamma_0$ is an analytical
function of $\lambda$, \Eq{CS12} is valid for all $\lambda$ rather
than for integer values only. Indeed, in the weak interaction
limit $|\lambda - 1| \ll 1$, \Eq{gamma0} yields $\gamma_0\approx
(\lambda - 1)/2$, in agreement with the corresponding limit of
\Eq{CS12}.

The behavior of the spectral function in the immediate vicinity of
the threshold, at $\overline{\xi_k}-\epsilon\lesssim k^2/\bar m$,
requires more  delicate consideration. Indeed, in this limit the velocity
of one of the quasiholes approaches
\begin{equation}
v_0(k) = -v\sign(k) + k/\bar m,
\label{CS14}
\end{equation}
while all other quasiholes have velocities $v_i\to \pm\,v$.
In other words, close to the threshold, almost all of the momentum
and energy are carried by a single quasihole,
\[
\overline{\xi_k} = - \frac{\bar m}{2}\, \bigl[v^2-
v_0^2(k)\bigr]\, .
\]
In order to evaluate the integral in \Eq{CS9}, we note that for
$\overline{\xi_k} -\epsilon\ll k^2/\bar m\,$ and
$\alpha=|k/p_F|\ll 1$ the conservation laws
[\,i.e., the $\delta-$functions in \Eq{CS9}] imply that for $k<0$
\[
X_1 \approx \alpha, \quad X_2\sim Y_1\sim \alpha^2.
\]
Actually, for $k<0$ it is just one of $x_i$, say, $x_{n_\lambda +
1}$, which is close to $\alpha$, while for $i=1,\ldots
n_\lambda$ one has $x_{i}\sim \alpha^2$. It is therefore
convenient to write
\[
x_{n_\lambda + 1} = \alpha + x_0, \quad x_0\sim\alpha^2,
\]
and introduce $\widetilde X_1 = X_1-\alpha =
\sum_{i=0}^{n_\lambda} x_i \sim\alpha^2$. In the leading (second)
order in $\alpha$ the $\delta-$functions in \Eq{CS9} can  be
approximated by
\[
\delta\!\left(\frac{\overline{\xi_k} - \epsilon}{\,2\bar m
v^2}\,-\widetilde X_1\right) \delta\!\left(\frac{\overline{\xi_k}
- \epsilon}{\,2\bar m v^2}\,-Y_1\right)
\]
[note that $0<(\overline{\xi_k} - \epsilon)\!/\bar m v^2\lesssim \alpha^2$].
At the same time, $x_{n_\lambda +1}$ in the form-factor should be
replaced by $\alpha$. With these approximations, the remaining
integrations are easily carried out resulting in
\begin{equation}
A_{k<0}(\epsilon)
\propto (\overline{\xi_k}-\epsilon)^{2\gamma_0^2-1},
\quad
0<\overline{\xi_k} -\epsilon\lesssim k^2\!/\bar m
\label{CS15}
\end{equation}
The fractional part of the exponent here is twice that in the
Luttinger liquid limit, see \Eq{CS11}. This is in agreement with the
leading logarithmic approximation result $A_{k<0}(\epsilon)\propto
(\overline{\xi_k}-\epsilon)^{\gamma_k^2-1}$, see \Eq{Ahole}. Indeed,
substitution of $V_k = -\lambda(\lambda - 1) \pi |k|/m$ into \Eq{mu}
gives $\mu_k = \mu_0 \approx \lambda-1$, independently of $k$ (this
agrees with the exact value\cite{MP_CSM} of the exponent that
governs the divergence of the structure factor, $\mu_0 =
1-1/\lambda$). \Eq{gammak} then yields $\gamma_k^2 = \mu_0^2/2
\approx 2\gamma_0^2$. Note also that the position of the spectral
edge $\overline{\xi_k}$, see \Eq{CS6}, in the limit $\lambda\to 1$
agrees with \Eq{xiover}.

Evaluation of $A_k(\epsilon)$ for $k>0$ proceeds similarly and
yields
\begin{equation}
A_{k>0}(\epsilon)\propto
(\overline{\xi_k}-\epsilon)^{\gamma_0^2 + \gamma_1^2}, \quad
0<\overline{\xi_k} -\epsilon\lesssim k^2\!/\bar m
\label{CS16}
\end{equation}
with $\gamma_1$ given by
\begin{equation}
\gamma_1^2=\frac{(\lambda -
3)^2}{4\lambda}\,.
\label{CS17}
\end{equation}
For a weak interaction
$\gamma_0^2 + \gamma_1^2 \approx 1-2\mu_0$, in agreement with
Eqs. \eq{2.40} and \eq{5.19}. Note that the exact exponent in \Eq{CS16} is
positive for any strength of interactions.

We have thus demonstrated that, at least in the hole region of the
spectrum $\epsilon<0$, the behavior expected for a generic 1D
system with a nonlinear dispersion is consistent with the exact
results obtained for the Calogero-Sutherland model. It would be
interesting to find a solvable model for which our conclusions for
the particle region of the spectrum can be similarly tested.

\section{Conclusions}
\label{conclusions}

The total number of particles and the total momentum are good
quantum numbers for an isolated homogeneous fermionic system,
regardless of its dimensionality and the interaction strength. In
higher dimensions ($D>1$) and for moderately strong interactions, an
excited state of the system with one extra particle and with
momentum $p$ is rather similar to the corresponding state of a free
Fermi gas. The similarity is encoded in the energy and momentum
dependence of the spectral function $A_p(\epsilon)=-(1/\pi)\im
G_p(\epsilon)$. The spectral function satisfies the exact sum rule,
\begin{equation}
\int\! d\epsilon\, A_p(\epsilon)=1\, .
\label{sumrule}
\end{equation}
In a Fermi liquid ($D>1$), the sum rule is almost completely
exhausted by the Lorentzian \Eq{FL}. The corresponding peak
is centered at the quasiparticle energy $\xi_p$ and has the width
$1/2\tau_p$ which decreases with the increase of the Fermi
energy, see \Eq{selfenergy}.
On the contrary, in a Luttinger liquid the spectral function, see \Eq{TL},
is manifestly non-Lorentzian. It diverges on the mass-shell $\epsilon=\xi_k$,
vanishes at $\epsilon<\xi_k$, and decays slowly with $\epsilon$ at
$\epsilon>\xi_k$. 

We demonstrated that for a nonlinear dispersion relation with
positive curvature, see \Eq{dispersion}, the domain where $A_k(\epsilon)$
differs from zero extends below the mass-shell, $\epsilon\geq \xi_k-k^2\!/m$.
The mass-shell $\epsilon=\xi_k$ now falls within the domain of continuous spectrum.
In this situation the Luttinger liquid result \Eq{TL} is no longer valid.
Instead of a one-sided power-law singularity at $\epsilon\to\xi_k$,
[cf. \Eq{TL}], the spectral function has a Lorentzian peak centered at
the mass-shell. The width of the peak $1/\tau_k$, see \Eq{1dselfenergy},
is much smaller than that in higher dimensions. In one dimension, the finite 
relaxation rate $1/\tau_k$ emerges only in the fourth order of the perturbation theory
in the interaction strength and increases with $k$ as $k^8$ (here $k$ is
momentum relative to $p_F$). 
At large enough momenta $k$, which are still exponentially small in the 
inverse interaction strength 
\begin{equation}
k \gtrsim \,p_{F}\,e^{-a/\gamma_0^2}\, ,
\label{k-star}
\end{equation} 
the Lorentzian peak in the spectral function carries most of the spectral 
weight. In other words,  for $k$ satisfying the inequality~\eq{k-star},
the dispersion nonlinearity restores the Fermi liquid character of the 
particle part of the spectrum. The numerical coefficient $a$ in the criterion 
\eq{k-star} depends on the fraction of the spectral weight chosen to fall 
within the Lorentzian.  

It should be emphasized that dispersion nonlinearity brings about
particle-hole asymmetry into the problem. Indeed, we found that
the nonlinearity does not affect qualitatively the hole region of the
spectrum $\epsilon<0$. The spectral function here resembles that
in the Luttinger liquid, although with the modified value of the exponent.

Finally, we mention that the emergence of a finite quasiparticle lifetime
brings some ramifications for the theory~\cite{Pustilnik2006} of the
structure factor $S(q,\omega)$ of an interacting 1D fermion system.
It smears the nonanalytical behavior of $S(q,\omega)$ at
$\omega=vq+q^2\!/2m$.
The smearing, however, affects only a tiny portion of the frequency
interval $|\omega-vq|\lesssim q^2\!/2m$. Indeed, the width of the smearing
region scales with $q$ as $1/\tau_q\propto q^8\!/m^3$, while the
width of the interval of interest is $q^2\!/2m$.

\begin{acknowledgments}
We thank Iddo Ussishkin for numerous discussions and for earlier
collaboration which stimulated this work. This project was supported
by DOE Grants DE-FG02-06ER46310 and DE-FG02-ER46311,
and by A.P. Sloan foundation.
\end{acknowledgments}

\begin{appendix}

\section{Vertex corrections}
\label{Sec:AppVerCor}

In this appendix we consider the dimensionless coupling vertex
$\Gamma$ characterizing the interaction of the $d$-particle with the
bosonic field $\phi$ describing excitations within the subbands $r$
and $l$, see \Eq{HdPotential}.  Our concern here is whether
corrections to $\Gamma$ due to interactions within the $d$-subband
affect the leading logarithmic series considered in \Sec{Sec:LogRenorm}.  
The vertex depends on four variables, 
$\Gamma = \Gamma(\epsilon,k,\omega,q)$. 
The dependence of $\Gamma$ on its arguments around the mass shell 
is not singular.  It is clear then that the relevant object is the on-shell 
vertex correction
\[
\delta\Gamma = \Gamma(\xi_k,k,\pm \,vq,q) - 1
\]
(the bare value of the vertex is $1$).

By construction of $r,l$-subbands (see \Sec{Sec:particlemassshell}),
the  momentum $q$ transferred to the bosonic field $\phi$ is restricted
to $|q|\leq \lambda k$. Accordingly, the relevant frequencies
$\omega=\pm \,vq$ satisfy $|\omega|\leq \lambda vk\ll\epsilon$.
It then follows that the two $d$-particle Green functions adjacent to the
bosonic field $\phi$ in the diagram such as that shown in
\Fig{Fig:Dressing}(b)
carry frequencies of the same sign, $\sign(\epsilon)=
\sign(\epsilon+\omega)$.
The on-shell vertex correction can then be expressed via the time-ordered
$d$-particle self-energy $\Sigma^T_k(\epsilon)$ as
\begin{equation}
\delta\Gamma =
\left.\frac{\partial
\Sigma_k^{T}}{\partial\epsilon}\right|_{\epsilon\to\xi_k}
+O(q)\,.
\label{A1}
\end{equation}
The latter is discussed in details in Sections \ref{Sec:PertTheor} and
\ref{Sec:LogRenorm} above.
In the second order in $d-d$ interaction the self-energy is purely real
and is given by
$\Sigma^{(2)}\sim\gamma_0^2(\epsilon-\xi_k)\ln(\Lambda/\lambda)$,
see \Sec{Sec:LogRenorm}. \Eq{A1} then yields
\begin{equation}
\re\delta \Gamma 
\sim \,\gamma_0^2
\ln(\Lambda/\lambda) \ll 1,
\label{A2}
\end{equation}
where we used \Eq{lambda}.
In the fourth order the self-energy acquires a finite imaginary part on the
mass-shell,
$\im\Sigma_k^{(4)}(\xi_k) = -1/2\tau_k$. Taking into account that
$\im\Sigma_k^{(4)}$ varies with $\epsilon$ on the scale $\sim k^2\!/m$, and
using \Eq{A1}, we find
\begin{equation}
\im \delta \Gamma 
\sim \frac{1/\tau_k}{k^2\!/m}\ll 1\,.
\label{A3}
\end{equation}

According to Eqs. \eq{A2} and \eq{A3}, vertex corrections
due to interactions within $d$-subband are small. As far as the summation
of the leading logarithmic contributions in \Sec{Sec:LogRenorm} is
concerned, all such corrections can be safely neglected.

\section{Derivation of Eq. \normalsize{\textbf{(\ref{FT})}}}
\label{Sec:AppDerivIntegr}

In this appendix we supply technical details needed to perform the
Fourier transform of \Eq{Greensolution} leading to
Eq.~\eqref{FT}. The coordinate integration can be performed at
once. With the help of the convolution theorem, the remaining integral 
over time be written as
\begin{equation}
G_k\left(\epsilon;\frac{\Lambda k^2}{mv},\Lambda k\right) = \int
\frac{ d \epsilon' }{ 2 \pi }\,
G_k^d(\epsilon - \epsilon')  {\cal F}_k(\epsilon') \, ,
                                                     \label{eq:App1010}
\end{equation}
where
\begin{equation}
G_k^d(\epsilon) = \frac{ 1 }{\epsilon - \xi_k + i / 2 \tau_k}
                                              \label{eq:App1014}
\end{equation}
is the bare $d$-particle Green function, see \Eq{gf-d},
and ${\cal F}_k(\epsilon)$ is given by
\begin{eqnarray}
{\cal F}_k(\epsilon) &= & \int\! d t\,  e^{ i  \epsilon  t } \left(
1 +  i \frac{\lambda k^2}{mv} ( v_d  +  v) t
\right)^{-\mu_+^2/4}
\nn\\
&&\quad \times \,\left( 1 - i \lambda k ( v_d  -  v) t
\right)^{-\mu_-^2/4},
\label{eq:App1018}
\end{eqnarray}
where we denoted $\mu_-=\mu_k$ and $\mu_+=\mu_{2p_F+k}$ 
for brevity. The integration in \Eq{eq:App1018} is carried out using 
the convolution theorem and the relation
\begin{equation}
 \int \frac{dt\,e^{i\epsilon t}}
{\left(1\pm i\kappa t\right)^{\mu^2_\pm/ 4}}
=
\frac{2\pi}{\Gamma (\mu_\pm^2/4)}
\frac{\theta(\pm\epsilon)}{|\epsilon|}
\left|\frac{\epsilon}{\kappa}\right|^{\,\mu_\pm^2/4}
e^{-|\epsilon|/\kappa},
\label{eq:App1020}
\end{equation}
where $\kappa \sim \lambda k^2/m$. Taking into account that 
$v_d - v \approx k/m$ and $v_d + v \approx 2 v $, we obtain
for $\mu_\pm \ll 1$   
\begin{equation}
{\cal F}_k(\epsilon)
= \frac{ \pi \mu_-^2 \mu_+^2}{8}\!
\int\! d\epsilon'\,
\frac{ \theta(\epsilon' )\theta( \epsilon' + \epsilon ) }{
\epsilon' (\epsilon' + \epsilon)} 
\left(\frac{\epsilon'}{\kappa } \right)^{\frac{\mu_-^2}{4}}\!
\left(\frac{\epsilon'  + \epsilon}{\kappa}\right)^{\frac{\mu_+^2}{4}}.
\label{eq:App1024}
\end{equation}
\Eq{eq:App1024} is valid for $|\epsilon | \lesssim \kappa
=\lambda k^2/m$. In the case $\epsilon > 0$ we have
\begin{equation}
{\cal F}_k(\epsilon)  = \frac{ \pi \mu_+^2 \mu_-^2 }{ 8
\epsilon }
\left(\frac{\epsilon }{\kappa} \right)^{\gamma_k^2} B\left(
\mu_-^2/4,1-\gamma_k^2\right) \, ,
\label{eq:App1032}
\end{equation}
where $4\gamma_k^2=\mu_-^2+\mu_+^2$, see \Eq{gammak}, and
$B$ is the standard beta-function. For weak interactions,
$\mu_\pm\ll 1$, one finds $B\approx 4/\mu_-^2$. The case $\epsilon
< 0$ is analyzed in a similar way. As a result, we obtain
\begin{equation}
{\cal F}_k(\epsilon) = \frac{  \pi }{ 2\, \epsilon }
\left|\frac{ \epsilon }{ \kappa } \right|^{\gamma_k^2}
\bigl[
\mu_+^2  \theta(\epsilon) +  \mu_-^2  \theta(- \epsilon)
\bigr]\, ,
\label{eq:App1053}
\end{equation}
covering both cases.
We now substitute Eqs.~\eqref{eq:App1014} and \eqref{eq:App1053}
into \Eq{eq:App1010}. Rewriting the integral in the latter
equation as the contour integral we find
\begin{widetext}
\begin{equation}
G_k\left(\epsilon;\frac{\Lambda k^2}{mv},\Lambda k\right) =
-\left( \lambda k^2/m \right)^{ - \gamma_k^2}\left[
\frac{ \mu_{2p_F+k}^2 }{4\gamma_k^2 }
\int_{C_+} \frac{ d z }{ 2
\pi i } \frac{ (-z)^{\gamma_k^2 - 1 } }{\epsilon - \xi_k + i /
2 \tau_k - z } -
\frac{ \mu_k^2 }{4\gamma_k^2 }
\int_{C_-} \frac{ d z }{ 2 \pi i }
\frac{ z^{\gamma_k^2 - 1 } }{\epsilon - \xi_k + i / 2 \tau_k -
z }
\right]\, ,
                                                    \label{eq:App1061}
\end{equation}
\end{widetext}
where the function $ z^{\alpha} $ is chosen to have a cut in the
complex plain along the negative real axis and contours $C_{\pm}$
run clockwise  around  the positive (negative) half of the real
axis. Closing the contours at infinity and taking the residue at
$z=\epsilon - \xi_k + i / 2 \tau_k$  we arrive at Eq.~\eqref{FT}.

\section{Derivation of Eq. \normalsize{\textbf{(\ref{5.33})}}}
\label{point_splitting}

Consider the operator
\begin{equation}
\Phi^\dagger (x) = \psi^{r\dagger}(-i\partial_x)\psi^{r\dagger}
\label{C1}
\end{equation}
entering \Eq{5.32}. Proceeding in a standard fashion, we subsitute here
$\psi^r=\sqrt{k_0}\,e^{i\varphi}$ with $\varphi\equiv\varphi^r$,
see \Eq{5.100}, and write the derivative as a finite difference,
\begin{equation}
\Phi^\dagger(x)
= \frac{-ik_0\,}{\,2\Delta} \,\,e^{-i\varphi(x)}\!
\left[e^{-i\varphi(x+\Delta)} - e^{-i\varphi(x-\Delta)}\right].
\label{C2}
\end{equation}
On making use of the identity
\[
e^a e^b =  \,\,:e^{a+b}\!:\,
e^{\left\langle ab + \frac{1}{2}\,(a^2 +b^2)\right\rangle}
\]
valid for any operators $a$ and $b$ which are \textit{linear}
in bosonic fields (the colons denote the normal ordering),
we find
\begin{eqnarray*}
e^{-i\varphi(x)}e^{-i\varphi(x\pm\Delta)}
&=& e^{-2\langle\varphi^2(0)\rangle}
:e^{-i\bigl[\varphi(x)+\varphi(x\pm\Delta)\bigr]}\!:
\nn\\
&&\qquad\qquad\qquad\times\,
\bigl(1\pm 2\pi i k_0 \Delta\bigr).
\end{eqnarray*}
We now approximate here
\[
:e^{-i\bigl[\varphi(x)+\varphi(x\pm\Delta)\bigr]}\!:
\,\approx\,\, :e^{-2i\varphi(x)}\!:\,\, = e^{2\langle\varphi^2(0)\rangle\,-\,2i\varphi(x)},
\]
so that
\begin{equation}
e^{-i\varphi(x)}e^{-i\varphi(x\pm\Delta)}
\approx e^{-2i\varphi(x)}\bigl(1\pm 2\pi i k_0 \Delta\bigr).
\label{C3}
\end{equation}
Eqs. \eq{C2} and \eq{C3} yield
\begin{equation}
\Phi^\dagger(x) = 2\pi k_0^2 \,e^{-2i\varphi(x)}.
\label{C4}
\end{equation}
Substituting this expression into
$\Psi^\dagger = \Phi^\dagger\psi^d$ [see Eqs. \eq{5.32} and \eq{C1}],
we arrive at \Eq{5.33}.

\end{appendix}


\end{document}